\documentclass{ws-ijmpd}
\usepackage{cite}
\usepackage{combelow}
\usepackage{bm}
\usepackage{slashed}
\usepackage{braket}
\usepackage{amsmath}
\usepackage{amsfonts}
\usepackage{mathrsfs}
\usepackage{url}

\def\draftnote{}
\def\trimmarks{}

\numberwithin{equation}{section}

\begin{document}
	
\markboth{Ambru\cb{s}, Kent and Winstanley}
{Analysis of scalar and fermion QFT on adS space-time}

%
%

\title{ANALYSIS OF SCALAR AND FERMION QUANTUM FIELD THEORY ON ANTI-DE SITTER SPACE-TIME}

\author{VICTOR E. AMBRU\cb{S}}

\address{Department of Physics, West University of Timi\cb{s}oara,\\
	Bd.~Vasile P\^arvan 4, Timi\cb{s}oara 300223, Romania\\
Victor.Ambrus@e-uvt.ro}

\author{CARL KENT}
\address{Consortium for Fundamental Physics, School of Mathematics and Statistics, \\ University of Sheffield,
	Hicks Building, Hounsfield Road, Sheffield. S3 7RH United Kingdom \\
C.Kent@sheffield.ac.uk}

\author{ELIZABETH WINSTANLEY}

\address{Consortium for Fundamental Physics, School of Mathematics and Statistics, \\ University of Sheffield,
	Hicks Building, Hounsfield Road, Sheffield. S3 7RH United Kingdom \\
E.Winstanley@sheffield.ac.uk}

\maketitle


\begin{abstract}
We study vacuum and thermal expectation values of quantum scalar and Dirac fermion fields on anti-de Sitter space-time. Anti-de Sitter space-time is maximally symmetric and this enables expressions for the scalar and fermion vacuum Feynman Green's functions to be derived in closed form.  We employ Hadamard renormalization to find the vacuum expectation values. The thermal Feynman Green's functions are constructed from the vacuum Feynman Green's functions using the imaginary time periodicity/anti-periodicity property for scalars/fermions.  Focussing on massless fields with either conformal or minimal coupling to the space-time curvature (these two cases being the same for fermions)  we compute the differences between the thermal and vacuum expectation values. We compare the resulting energy densities, pressures and pressure deviators with the corresponding classical quantities calculated using relativistic kinetic theory.
\end{abstract}




\section{Introduction}
\label{sec:intro}

Quantum field theory (QFT) on curved space-time is a semi-classical approximation to quantum gravity, in which quantum fields propagate on a fixed, classical, background geometry. 
Within this approach, a key quantity is the renormalized stress-energy tensor (SET) $\braket{{\hat {T}}_{\mu \nu }}$ which governs the back-reaction of the quantum field on the space-time geometry through the semi-classical Einstein equations
\begin{equation}
	G_{\mu \nu } + \Lambda g_{\mu \nu } = 8\pi \braket{{\hat {T}}_{\mu \nu }}.
	\label{eq:SCEE}
\end{equation}
We employ units in which $k_{B}=G=c=\hbar = 1$ and our metric signature is $(-,+,+,+)$.

In this report we study quantum scalar and Dirac fermion fields on anti-de Sitter (adS) space-time.  This is motivated by the adS/CFT (conformal field theory) correspondence (see, for example, \cite{Aharony:1999ti} for a review), according to which quantum gravity on certain asymptotically adS space-times is dual to a conformal field theory on the space-time boundary.  An understanding of the behaviour of quantum fields on the simplest asymptotically adS space-time, namely adS itself, may provide insights into the semi-classical approximation to quantum gravity on more complicated asymptotically adS space-times, such as those containing black holes.

One advantage of working on pure adS space-time is that the maximal symmetry enables many quantities relevant to QFT to be derived in closed form. Here we focus on the vacuum and thermal expectation values of the SET for scalar and Dirac fermion fields. Two of us have previously studied in depth the properties of vacuum \cite{Ambrus:2015mfa}  and thermal \cite{Ambrus:2017cow} states for Dirac fermions on adS (see also \cite{Ambrus:2017vlf,Ambrus:2014fka} for some complementary discussion).
For scalar fields, the vacuum state is considered on $n$-dimensional adS in \cite{Kent:2014nya}, and work on the corresponding thermal states is ongoing \cite{KentEW} (we present some preliminary results here).
Our focus in this report is to compare the properties of quantum scalar and Dirac fermion fields on four-dimensional adS space-time. In the Dirac fermion case, it has proved to be insightful to study thermal states not only within the framework of QFT, but also in classical relativistic kinetic theory (RKT) \cite{Ambrus:2017cow, Ambrus:2017vlf}.  Here we extend the RKT results for Dirac fermions presented in \cite{Ambrus:2017cow} to scalar particles, and compare with those arising in QFT.  

The outline of this report is as follows. In Sec.~\ref{sec:Minkowski} we briefly review some key features of QFT and RKT on Minkowski space-time, for both scalar and Dirac fermion particles. Our study of adS space-time begins in Sec.~\ref{sec:adS} with an outline of some aspects of the geometry of adS space-time which are particularly relevant for QFT on this background. For comparison with later QFT results, in Sec.~\ref{sec:RKT} we discuss the properties of thermal scalars and Dirac fermions in classical RKT on adS space-time.   Our main QFT results are in Secs.~\ref{sec:scalar} and \ref{sec:fermion}, where we consider the vacuum expectation values (v.e.v.s) and thermal expectation values (t.e.v.s) of the SET for scalar and Dirac fermion fields respectively on adS space-time.
Sec.~\ref{sec:conc} contains some brief conclusions. For the remainder of this report, we shall use the terminology ``fermion'' to mean ``Dirac fermion'' as we shall not consider other types of fermion field.

\section{QFT and RKT on Minkowski space-time}
\label{sec:Minkowski}

We begin with a brief review of the salient features of QFT and RKT on Minkowski space-time, which will be useful for comparing with the adS case. 

\begin{figure}
	\begin{center}
	\includegraphics[width=6cm]{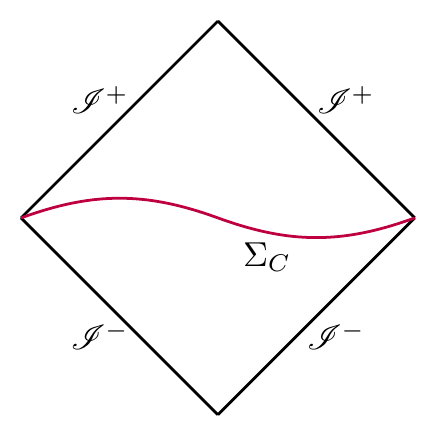}
	\end{center}
	\caption{Conformal diagram of Minkowski space-time with a Cauchy surface $\Sigma _{C} $. ${\mathscr{I}}^{\pm }$ are future and past null infinity. \label{fig:Minkowski}}
\end{figure}

Minkowski space-time, as well as having zero curvature, is maximally symmetric and globally hyperbolic, which greatly simplifies the quantum theory of free fields on this background.
The fact that Minkowski space-time is globally hyperbolic means that it possesses a Cauchy surface $\Sigma _{C} $ (see Fig.~\ref{fig:Minkowski}).  For a scalar or fermion field, specifying suitable initial data on $\Sigma _{C}$ uniquely determines the evolution of the field throughout the space-time.

Due to the maximal symmetry of Minkowski space-time, the vacuum Feynman Green's function for both scalar and fermion fields depends only on the geodetic interval $s_{{\rm {M}}}(x,x')$ between the space-time points $x=(t,{\bm {x}})$ and $x'=(t',{\bm {x'}})$, which is given by 
\begin{equation}
s_{{\rm {M}}}^{2} = -\Big( x^{\hat {\alpha }}- x^{\hat {\alpha }'}\Big) \Big( x_{\hat {\alpha }} - x_{\hat {\alpha }'}\Big) ,
\label{eq:sMinkowski}
\end{equation}
where we have denoted Minkowski space-time indices using hats for ease of comparison with our later results on adS space-time. Indices with a prime denote quantities at the space-time point $x'$.
The vacuum Feynman Green's function for a scalar field of mass $m$ is given, in terms of $s_{\rm{M}}$, by \cite{Birrell:1982ix}
\begin{subequations}
\begin{equation}
iG_{\rm {vac}}^{\rm {M}} (x,x') = \frac{im}{8\pi s_{\rm{M}}} H_{1}^{(2)} (ms_{{\rm {M}}})
\label{eq:scalarvacM}
\end{equation}
where $H_{1}^{(2)}$ is a Hankel function of the second kind.
The vacuum Feynman Green's function for a fermion field of mass $m$ can be written in terms of the scalar vacuum Feynman Green's function (\ref{eq:scalarvacM}) as follows \cite{Muck:1999mh}
\begin{equation}
iS_{\rm {vac}}^{\rm {M}} (x,x') = \left( i{\slashed{\partial}} + m\right) iG_{\rm {vac}}^{\rm {M}}(x,x'), 
\label{eq:fermionvacM}
\end{equation}
\end{subequations}
where $\slashed{\partial} = \gamma ^{\hat {\alpha }}\partial _{\hat {\alpha }}$ and $\gamma ^{{\hat {\alpha }}}$ are the usual Dirac matrices.
The corresponding thermal Feynman Green's functions at inverse temperature $\beta $ are constructed from the above vacuum Feynman Green's functions by imposing appropriate periodicities in imaginary time \cite{Birrell:1982ix}:
\begin{subequations}
	\label{eq:thermal_M}
\begin{eqnarray}
G_{\beta }^{\rm {M}}(x,x') & = & 
\sum _{j=-\infty }^{\infty } G_{\rm {vac}}^{\rm {M}} (t+ij\beta ,{\bm {x}}; t',\bm {x'}) ,
\label{eq:scalar_thermal_M}
\\
S_{\beta }^{\rm {M}}(x,x') & = & 
\sum _{j=-\infty }^{\infty }(-1)^{j} S_{\rm {vac}}^{\rm {M}} (t+ij\beta ,{\bm {x}}; t',\bm {x'}).
\label{eq:fermion_thermal_M} 
\end{eqnarray}
\end{subequations}
V.e.v.s and t.e.v.s of the SET are computed from the corresponding Feynman Green's functions. 
For a quantum scalar field, the relevant expression is (where $\eta _{\hat {\alpha} \hat {\nu}}$ is the Minkowski metric) \cite{Birrell:1982ix}:
\begin{subequations}
	\label{eq:SET_M}
\begin{equation}
\braket{T_{\hat{\alpha}\hat{\nu}}}^{{\mathrm {S}},\mathrm {M}}_{\mathrm {vac}/\beta }
= \lim_{x'\rightarrow x} 
\left\{ \left[
\delta _{\hat{\nu }}^{\hat {\nu }'}\partial _{\hat {\alpha }} \partial _{\hat {\nu }'} -\frac{1}{2}\eta _{\hat {\alpha }{\hat {\nu}}}\eta ^{\hat {\sigma }\hat {\lambda }'} \partial _{{\hat {\sigma }}}\partial _{{\hat {\lambda }}'} - \frac{1}{2} \eta _{{\hat {\alpha}}\hat {\nu}} m^{2} 
\right] iG^{M}_{\rm {vac}/\beta }
\right\} ,
\label{eq:scalar_set_M}
\end{equation}
while for a quantum fermion field, we have \cite{Birrell:1982ix}
\begin{equation}
\braket{T_{\hat{\alpha}\hat{\nu}}}^{\mathrm {F},\mathrm {M}}_{{\rm {vac}}/\beta } = \frac{i}{2} \lim_{x'\rightarrow x} {\mathrm {Tr}}\left\{
\left[\gamma_{(\hat{\alpha}} \partial _{\hat{\nu})} iS^{\mathrm {M}}_{{\rm {vac}}/ \beta }(x, x') -
\partial _{\hat{\nu}'} [iS^{\mathrm {M}}_{{\rm {vac}}/\beta }(x,x')]
\gamma_{(\hat{\alpha}} \delta  _{\hat {\nu })}^{{\hat {\nu}}'}
\right]  \right\}.
\label{eq:fermion_set_M}
\end{equation} 
\end{subequations}
Both the v.e.v.~and the t.e.v.~are infinite when the limit $x'\rightarrow x$ is taken.
On Minkowski space-time, this is resolved by setting the v.e.v.~of the SET to be equal to zero. For the t.e.v.s, this means that the vacuum contributions $G_{\rm {vac}}^{{\mathrm {M}}}(x,x')$, $S_{\rm {vac}}^{\rm {M}}(x,x')$ (corresponding to $j=0$)  are subtracted from the sums in (\ref{eq:thermal_M}).
We are effectively then considering the differences between the t.e.v.s and the v.e.v.s, which we denote by 
\begin{equation}
\braket{:T_{\hat{\alpha}\hat{\nu}}:}_{\beta }^{\mathrm {M}} \equiv
\braket{T_{\hat{\alpha}\hat{\nu}}}_{\beta }^{\mathrm {M}} -
\braket{T_{\hat{\alpha}\hat{\nu}} }_{\rm {vac}}^{\mathrm {M}} 
\end{equation}
for both scalar and fermion fields.
When the limit $x'\rightarrow x$ is taken in (\ref{eq:SET_M}), the vacuum Feynman Green's functions appearing in (\ref{eq:thermal_M}) depend on the geodetic interval $s_{\mathrm {M}}$ (\ref{eq:sMinkowski}) between the points $(t,\bm {x})$ and $(t+ij\beta ,\bm{x})$, which takes the form
\begin{equation}
s_{j} =-i \beta \left|j \right| .
\end{equation}
Using the results
\begin{equation}
H_{1}^{(2)}(ms_{j}) = -\frac{2}{\pi }K_{1} (m\beta \left|j \right|) , \qquad
H_{2}^{(2)}(ms_{j}) = -\frac{2i}{\pi} K_2(m\beta\left|j\right|)
\end{equation}
where $K_{u}$ is a modified Bessel function, we find that the t.e.v.s of the SET for a quantum scalar and fermion field both take the perfect fluid form
\begin{equation}
\braket{:T^{\hat{\alpha}\hat{\nu}}:}_{\beta }^{{\mathrm {S/F}},\mathrm {M}} 
= {\mathrm {Diag}} \left\{ E^{\mathrm{S/F}}_{{\mathrm {M}}}(\beta ) , P^{\mathrm {S/F}}_{{\mathrm {M}}}(\beta ), P^{\mathrm {S/F}}_{{\mathrm {M}}}(\beta ),
P^{\mathrm {S/F}}_{{\mathrm {M}}}(\beta ) \right\} ,
\label{eq:fluid}
\end{equation}
where, for a quantum scalar field, the energy density $E_{\mathrm {M}}^{S}(\beta )$ and pressure $P_{\mathrm {M}}^{S}(\beta )$ are given by 
\begin{subequations}
	\label{eq:RKT_M}
\begin{equation}
E_{\mathrm {M}}^{S}(\beta ) - 3P_{\mathrm {M}}^{S}(\beta ) = \frac{m^{3}}{2\pi ^{2}\beta } \sum _{j=1}^{\infty } \frac{1}{j} K_{1}(mj\beta ) ,
\quad
P_{\mathrm {M}}^{S}(\beta ) = \frac{m^{2}}{2\pi ^{2}\beta ^{2}} \sum _{j=1}^{\infty } \frac{1}{j^{2}} K_{2}(mj\beta ),
\end{equation}	
while the corresponding expressions for a quantum fermion field are
\begin{eqnarray}
E_{\mathrm {M}}^{F}(\beta ) - 3P_{\mathrm {M}}^{F}(\beta ) & = & \frac{2m^{3}}{\pi ^{2}\beta } \sum _{j=1}^{\infty } \frac{\left(-1\right) ^{j-1}}{j} K_{1}(mj\beta ) ,
\nonumber \\
P_{\mathrm {M}}^{F}(\beta ) & = & \frac{2m^{2}}{\pi ^{2}\beta ^{2}} \sum _{j=1}^{\infty } \frac{\left(-1\right) ^{j-1}}{j^{2}} K_{2}(mj\beta ).
\end{eqnarray}	
	\end{subequations}
Simple closed-form expressions can be found in the massless limit $m\rightarrow 0$ using the asymptotic properties of Bessel functions:
\begin{equation}
\left. E_{\mathrm {M}}^{S}(\beta )\right\rfloor _{m=0}  =  \left. 3P_{\mathrm {M}}^{S}(\beta )\right\rfloor _{m=0} =\frac{\pi ^{2}}{30\beta^4} ,
\qquad
\left. E_{\mathrm {M}}^{F}(\beta )\right\rfloor _{m=0}  =  \left. 3P_{\mathrm {M}}^{F}(\beta )\right\rfloor _{m=0} =\frac{7\pi ^{2}}{60 \beta^4} .
\label{eq:massless_M}
\end{equation}
For all field masses, both the energy density and pressure are constants throughout Minkowski space-time. Minkowski space-time is maximally symmetric (with no space-time point preferred) and considering a quantum state at a temperature $\beta ^{-1}$ does not break this spatial invariance. Although the definition of the thermal Feynman Green's functions (\ref{eq:thermal_M}) involves a choice of time $t$, this procedure is invariant under time translations.  
Indeed, the generators of 
time and space translations both commute with 
the density operator corresponding to the thermal state considered in
this section \cite{Becattini:2014yxa}.
In contrast, if one considers a rigidly-rotating thermal state on Minkowski space-time, the Hamiltonian corresponding to the time coordinate of a rigidly-rotating observer does not commute with all generators of spatial translations. In this case the components of the t.e.v.~of the SET pick up a dependence on the distance from the rotation axis \cite{Ambrus:2014uqa}.

For comparison with our later work on adS, it is instructive to compare the above QFT results on Minkowski space-time with the classical energy density and pressure for a thermal gas of particles, computed in the framework of RKT.
For particles with mass $m$, momentum $p^{\hat {\alpha }}$, at inverse temperature $\beta $, the Bose-Einstein and Fermi-Dirac thermal distribution functions are, respectively \cite{Ambrus:2016ocv,Ambrus:2015rnx,Florkowski:2014sda}
\begin{subequations}
	\label{eq:thermalf}
\begin{eqnarray}
f^{\rm {S}}_{\beta } & = & \frac{Z}{8\pi ^{3} \left( e^{\beta p^{\hat{0}} } - 1 \right)}
 = 
\frac {1}{8 \pi^{3}}
\sum_{j = 1}^\infty 
e^{-j\beta p^{\hat{0}}}, 
\label{eq:feq_M_S}
\\
f^{\rm {F}}_{\beta } & = & \frac{Z}{8\pi ^{3} \left( e^{\beta p^{\hat{0}} } + 1 \right)}
=
\frac {1}{2 \pi^{3}}
\sum_{j = 1}^\infty (-1)^{j-1}
e^{-j\beta p^{\hat{0}}},
\label{eq:feq_M_F}
\end{eqnarray}
\end{subequations}
where we consider a state in which the fluid is at rest and have assumed that the chemical potential vanishes. For scalars, the number of degrees of freedom per particle is $Z=1$, while for fermions it is $Z=4$.
Integrating the distribution function with respect to the particle momentum gives the classical thermal SET:
\begin{equation}
T^{\hat{\alpha}\hat{\nu}\,{\mathrm {S/F}}}_{\rm {RKT}} = \int \frac{d^3p}{p^{\hat{0}}} f^{\mathrm {S/F}}_{\beta }\, p^{\hat{\alpha }} p^{\hat {\nu}}.
\end{equation}
Remarkably, on Minkowski space-time, for both scalars and fermions the classical thermal SET is identical to the quantum t.e.v.~(\ref{eq:fluid}) \cite{Itzykson:1980rh,Ambrus:2015rnx,Ambrus:2016ocv}:
\begin{equation}
T^{\hat{\alpha}\hat{\nu}\,{\mathrm {S/F}}}_{\rm {RKT}} =
{\mathrm {Diag}} \left\{ E^{\mathrm{S/F}}_{{\mathrm {M}}}(\beta ) , P^{\mathrm {S/F}}_{{\mathrm {M}}}(\beta ), P^{\mathrm {S/F}}_{{\mathrm {M}}}(\beta ),
P^{\mathrm {S/F}}_{{\mathrm {M}}}(\beta ) \right\} .
\end{equation}
This means that, on Minkowski space-time, there are no quantum corrections to the thermal SET.

\section{AdS space-time}
\label{sec:adS}

Before we examine the behaviour of quantum scalar and fermion fields on adS space-time, we now review some of the key features of the geometry of adS space-time which are relevant for QFT on this background. Other aspects of adS space-time geometry are discussed in, for example, \cite{Hawking:1973uf,Bengtsson:1998,Moschella:2005,Socolovsky:2017nff}.
We also discuss the need to impose boundary conditions on classical fields propagating on this background.

\subsection{Geometry of adS space-time}
\label{sec:adSgeometry}

\begin{figure}
\begin{center}
	\begin{tabular}{cc}
		adS & CadS\\
\includegraphics[width=6cm]{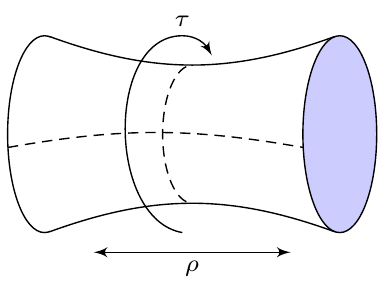} & 
\includegraphics[width=6cm]{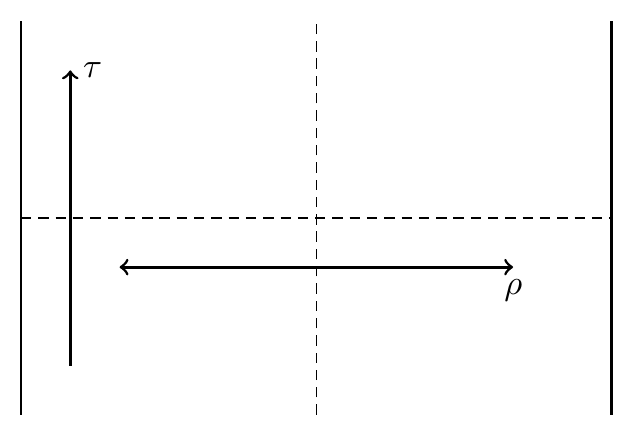}
\\
$-\pi < \tau \le \pi $  & $-\infty < \tau <\infty $ 
\end{tabular}
\end{center}
\caption{AdS (left) and CadS (right) space-times with two dimensions suppressed. The dashed lines are examples of constant $\tau $ and $\rho $ curves. The two ends of the hyperboloid in the left-hand diagram are directed towards spatial infinity for $\theta =0$ and $\theta =\pi $.
	\label{fig:adShyp}}
\end{figure}

Four-dimensional adS space-time is defined as the hyperboloid
\begin{equation}
-x_{0}^{2}+x_{1}^{2}+x_{2}^{2}+x_{3}^{2}-x_{4}^{2} = -a^{2}
\end{equation}
with radius of curvature $a$, embedded in five-dimensional flat space-time with two time directions, whose metric is
\begin{equation}
ds^{2}  =  -dx_{0}^{2} + dx_{1}^{2}+dx_{2}^{2}
+dx_{3}^{2}-dx_{4}^{2}.
\end{equation}
Defining the dimensionless intrinsic coordinates $(\tau, \rho, \theta , \varphi)$ via
\begin{equation}
\begin{array}{rclrclrcl}
x_{0} &  = & a \cos \tau \sec \rho , &
\quad x_{1} & = & a \tan \rho \cos \theta , &
\quad x_{2} & = & a \tan \rho \sin \theta \cos \varphi ,  \\
x_{3} & = & a\tan \rho \sin \theta \sin \varphi , &
\quad x_{4} & = & a \sin \tau \sec \rho , & & 
\end{array}
\end{equation}
where $-\pi < \tau \le \pi $ (with $\tau=-\pi$ and $\tau=\pi$ identified) and $0\le \rho < \pi /2$, the metric for adS space-time is 
\begin{equation}
ds^{2}_{\mathrm {adS}} = a^{2} \sec ^{2}\rho \left[ -d\tau ^{2}+ d \rho ^{2} + \sin ^{2} \rho \left( d\theta ^{2} + \sin ^{2}\theta \, d\varphi ^{2} \right) \right] .
\label{eq:adSmetric}
\end{equation}
The space-like coordinate $\rho $ runs perpendicular to the waist of the hyperboloid (see Fig.~\ref{fig:adShyp}), with $\rho =\pi /2$ corresponding to the adS space-time boundary. The time-like coordinate $\tau $ runs parallel to the waist of the hyperboloid, as shown in Fig.~\ref{fig:adShyp}. This means that adS possesses unphysical closed time-like curves.  To avoid these, we can unwrap the hyperboloid and pass to the covering space of adS (CadS), in which the range of the coordinate $\tau $ is extended to infinity. 

AdS is a maximally symmetric space-time with constant negative curvature. This can be seen in the form of the Riemann tensor
\begin{equation}
	R_{\mu \nu \lambda \sigma } =- \frac{1}{a^{2}} \left( g_{\mu \lambda } g_{\nu \sigma } -
g_{\mu \sigma} g_{\nu \lambda }\right) ,
\end{equation}
from which the Ricci scalar is
\begin{equation}
R = -\frac{12}{a^{2}}.
\label{eq:Ricci}
\end{equation}

\begin{figure}
	\begin{center}
		\includegraphics[width=7cm]{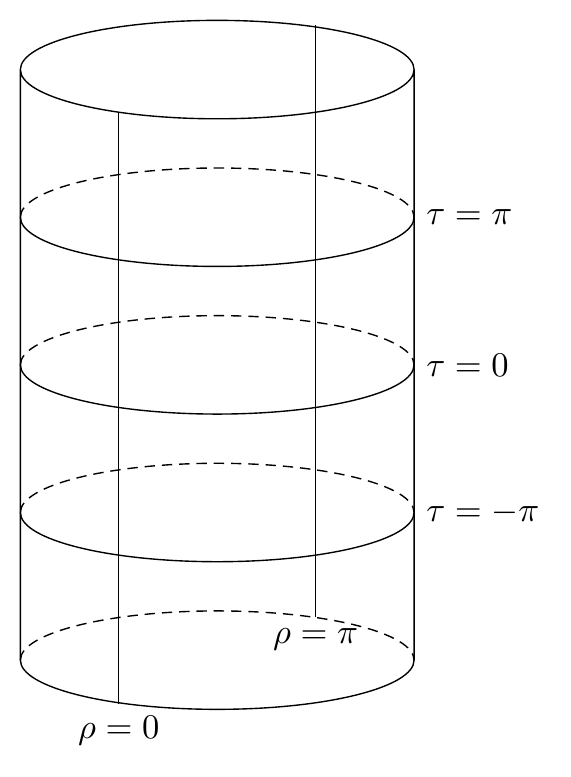}
	\end{center}
\caption{ESU with two dimensions suppressed. The cylinder extends to infinity in both directions. \label{fig:esu}}
\end{figure}

\begin{figure}
	\begin{center}
		\includegraphics[width=6cm]{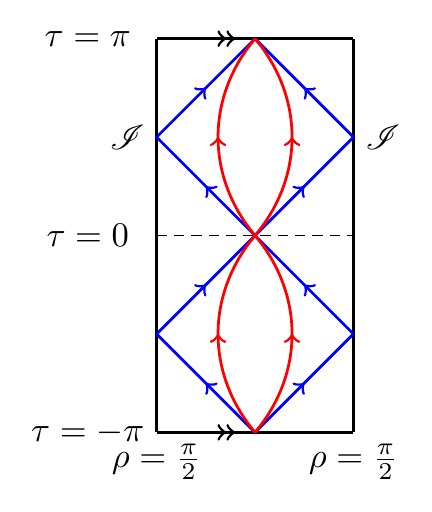}
		\includegraphics[width=6cm]{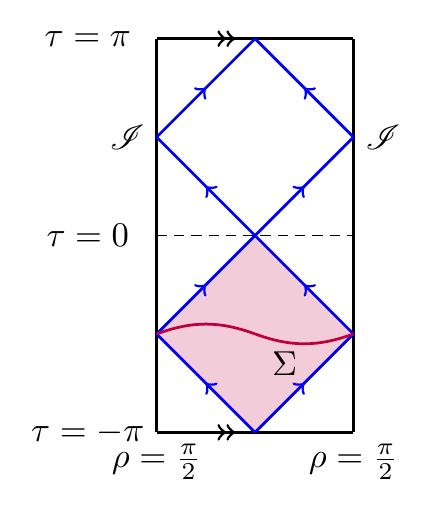}
	\end{center}
	\caption{The conformal diagram for adS space-time, with null infinity denoted by ${\mathscr{I}}$. 
		The horizontal lines marked with double arrows are to be identified.
		In the left-hand figure, some sample time-like geodesics are shown as red curved lines, and some sample null geodesics are shown as blue diagonal lines. 
		The right-hand figure shows a space-like surface $\Sigma $ and the shaded region is the region of the space-time  on which the evolution of a field is determined by initial data prescribed on $\Sigma $.
		\label{fig:adSconf}}
\end{figure}

From the adS metric (\ref{eq:adSmetric}) with $0\le \rho < \pi /2$, it is clear that adS space-time is conformal to (a part of) the Einstein static universe (ESU), which has metric
\begin{equation}
ds^{2}_{\rm {ESU}}  =   -d\tau ^{2}+ d \rho ^{2} 
+ \sin ^{2} \rho \left( d\theta ^{2} + \sin ^{2}\theta \, d\varphi ^{2} \right) .
\end{equation}
The ESU, as shown in Fig.~\ref{fig:esu}, is a cylinder and the radial 
coordinate $\rho $ has the range $0\le \rho < \pi $.
In contrast, for CadS, the coordinate $\rho $ has the range $0\le \rho < \pi /2$, and accordingly CadS is conformal to one half of the ESU.
For adS, the conformal diagram is shown in Fig.~\ref{fig:adSconf}, where the horizontal lines marked with double arrows are to be identified.

\subsection{Boundary conditions for classical fields on adS}
\label{sec:boundary}

From the conformal diagram in Fig.~\ref{fig:adSconf}, it can be seen that null infinity ${\mathscr{I}}$ is time-like. 
While ${\mathscr{I}}$ is an infinite proper distance away from the origin at $\rho =0$, and accordingly cannot be reached by time-like geodesics (red curved lines in Fig.~\ref{fig:adSconf}), null geodesics reach ${\mathscr{I}}$ in finite affine parameter (blue diagonal lines in Fig.~\ref{fig:adSconf}).
As a result, neither adS nor CadS are globally hyperbolic space-times. To see this, consider the space-like hypersurface $\Sigma $ shown in Fig.~\ref{fig:adSconf}.
The red shaded area in Fig.~\ref{fig:adSconf} represents the region on which the evolution of a classical scalar or fermion field is determined by initial data on $\Sigma $.
This region is a diamond bounded by null geodesics which reach $\mathscr{I}$. Hence $\Sigma $ cannot be a Cauchy surface in either adS or CadS.

\begin{figure}
	\begin{center}
		\begin{tabular}{cc}
			Transparent boundary conditions & Reflective boundary conditions \\
		\includegraphics[width=6cm]{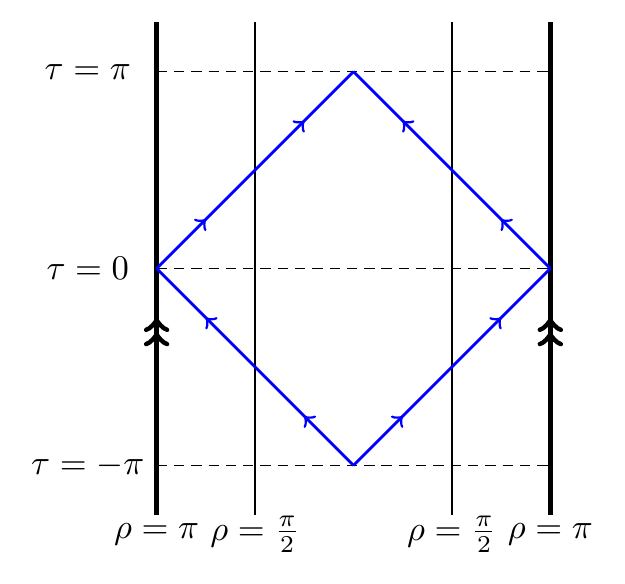} &
		\includegraphics[width=6cm]{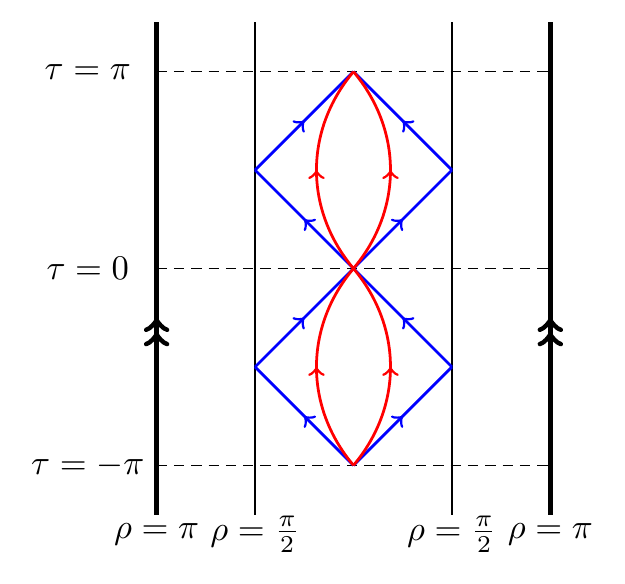}
	\end{tabular}
	\end{center}
\caption{Boundary conditions for a classical field on adS. The diagrams show the ESU from Fig.~\ref{fig:esu}, with the cylinder opened up, so that the vertical lines marked with double arrows should be identified. AdS space-time lies between the interior vertical lines.  The blue diagonal lines denote null geodesics (on the ESU in the left-hand figure and on adS in the right-hand figure). The red curved lines in the right-hand figure are time-like geodesics. \label{fig:adSBCs}}
\end{figure}

Boundary conditions on a classical field therefore need to be applied on ${\mathscr{I}}$ \cite{Avis:1977yn} in order for initial data on $\Sigma $ to determine the evolution of the field on the entire adS space-time, which is necessary for a well-defined quantum field theory. 
While adS is not globally hyperbolic, the ESU is, and appropriate boundary conditions can be constructed by considering the embedding of adS in the ESU, as shown in Fig.~\ref{fig:adSBCs}.

Suppose first of all that we have a conformally invariant classical field (for example, a massless fermion field). Since adS and the ESU are conformally related, we can consider the evolution of the field on the ESU, and then restrict to adS to give the evolution of the field on adS. This corresponds to transparent boundary conditions (left-hand diagram in Fig.~\ref{fig:adSBCs}) \cite{Avis:1977yn}.  Since conformally invariant fields are by definition massless (a field mass would introduce a length scale) and it is only null geodesics that reach ${\mathscr{I}}$ on adS, transparent boundary conditions can only be applied to conformally invariant fields. For this reason we will not consider transparent boundary conditions further. 

The second possibility is to regard adS as a box embedded in the ESU, and impose reflective boundary conditions on the surface of the box, in an analogous way to the procedure for the quantization of fields within a box in Minkowski space-time (right-hand diagram in Fig.~\ref{fig:adSBCs})  \cite{Avis:1977yn}.  Since time-like geodesics do not reach ${\mathscr {I}}$ in adS, reflective boundary conditions can be applied to all classical fields, whether or not they are conformally invariant.
For this reason, we will consider only reflective boundary conditions for the remainder of this report. Although in this section we have discussed boundary conditions for fields on adS, the same considerations apply to fields on CadS.  For the rest of this report, we consider fields on CadS rather than adS, and the abbreviation ``adS'' should be taken to mean ``CadS''.

\subsection{Useful quantities}
\label{sec:useful}

In the following sections, we turn to RKT and QFT on the adS space-time background. To aid comparison with our Minkowski space-time results from Sec.~\ref{sec:Minkowski}, we will write the SET components relative to the following  tetrad:
\begin{equation}
\begin{array}{rclrclrclrcl}
\omega^{\hat {\tau }}& =& {\displaystyle{\frac{a \, d\tau}{\cos\rho}}} , &
\omega^{\hat {\rho }} & =& {\displaystyle{\frac{a  \, d\rho}{\cos\rho} }}, &
\omega^{\hat {\theta }} & =& {\displaystyle{\frac{a \, d\theta}{\cot\rho} }}, &
\omega^{\hat {\varphi }}  &=& {\displaystyle {\frac{a \sin\theta}{\cot\rho} d\varphi }}, \\
e_{\hat {\tau }} & =& {\displaystyle { \frac{1}{a} \cos\rho \, \partial_\tau }}, &
e_{\hat {\rho}}  & =& {\displaystyle {\frac{1}{a} \cos\rho  \, \partial_\rho }}, &
e_{{\hat {\theta }}} &=& {\displaystyle {\frac{1}{a} \cot\rho \, \partial_{\theta} }}, &
e_{{\hat {\varphi }}} & =& {\displaystyle { \frac{1}{a} \frac{\cot\rho}{\sin\theta} \partial_{\varphi } }}.
\end{array}
\label{eq:frame}
\end{equation}
Since the boundary of adS is an infinite proper distance away from the origin at $\rho =0$, we will find it helpful in later sections to plot various quantities as functions of the dimensionless geodetic distance $\mu _{{\mathrm {adS}}}$ between the origin and a point with radial coordinate $\rho $:
\begin{equation}
\mu _{\mathrm {adS}}(\rho ) = \frac{1}{a} \int _{0}^{\rho } ds _{\mathrm {adS}} =  \cosh ^{-1} \left( \sec \rho \right) . 
\label{eq:muadS}
\end{equation}

\section{Relativistic kinetic theory on adS space-time}
\label{sec:RKT}

Before we study QFT for scalars and fermions on adS space-time, in this section we consider  classical RKT on this background.  
Since adS is a curved space-time, the inverse temperature $\beta $ in the thermal distribution functions (\ref{eq:thermalf}) must be replaced by the local inverse temperature ${\tilde {\beta }}$, which is given by \cite{Ambrus:2016ocv,tolman30}
\begin{equation}
{\tilde {\beta }} = \frac {\beta }{a}\sqrt{-g_{\tau \tau }} =
\frac{\beta}{\cos \rho},
\label{eq:tolman}
\end{equation}
where  $g_{\tau \tau }$ is the $(\tau, \tau )$ component of the adS metric (\ref{eq:adSmetric}) and $\beta =  {\tilde {\beta }}(\rho = 0)$ is the local inverse temperature at the coordinate origin.
The local inverse temperature ${\tilde {\beta }}$ (\ref{eq:tolman}) is not constant in space, and accordingly we expect that the energy density and pressure of the thermal gas of particles will also not be constant, unlike the situation in Minkowski space-time. In particular, we  note that ${\tilde {\beta }}\rightarrow \infty $ (and hence the local temperature vanishes) as $\rho \rightarrow \pi/2$ and the adS boundary is approached.

Although adS, like Minkowski space-time, is maximally symmetric, the local inverse temperature (\ref{eq:tolman}) is not invariant under spatial translations. 
It is clear from (\ref{eq:tolman}) that specifying a temperature corresponds not only to a choice of time coordinate $\tau $, but in addition a preferred origin has been selected as the point where the local inverse temperature is $\beta $ \cite{Allen:1986ty}.
Therefore the introduction of a temperature breaks the maximal  $SO(2,3)$ symmetry of adS space-time down to $SO(2)\times SO(3)$ \cite{Allen:1986ty}.
Furthermore, in contrast to the situation on Minkowski space-time, the generators of space translations on adS do not commute with the density operator corresponding to a global thermal state \cite{Cotaescu:2017ywe,Cotaescu:2017iha}.

The computation of the energy density $E_{\rm{adS}}^{\mathrm{S/F}}(\beta )$ and pressure $P_{\mathrm {adS}}^{\mathrm {S/F}}(\beta )$ for scalars and fermions on adS follows that in Sec.~\ref{sec:Minkowski} on Minkowski space-time, but with $\beta $ replaced by ${\tilde{\beta }}$ (\ref{eq:tolman}).
We therefore find, for scalars:
\begin{subequations}
	\label{eq:adSRKT}
	\begin{eqnarray}
	E_{\mathrm {adS}}^{S,{\mathrm {RKT}}}(\beta ) - 3P_{\mathrm {adS}}^{S,\mathrm{RKT}}(\beta ) & = & \frac{m^{3}\cos \rho }{2\pi ^{2}\beta } \sum _{j=1}^{\infty } \frac{1}{j} K_{1}\left(\frac{mj\beta }{\cos \rho }\right) ,
	\nonumber \\
	P_{\mathrm {adS}}^{S,{\mathrm{RKT}}}(\beta ) & = & \frac{m^{2}\cos ^{2}\rho }{2\pi ^{2}\beta ^{2}} \sum _{j=1}^{\infty } \frac{1}{j^{2}} K_{2}\left(\frac{mj\beta }{\cos \rho }\right) ,
	\label{eq:scalarRKT}
	\end{eqnarray}	
	while the corresponding expressions for fermions are:
	\begin{eqnarray}
	E_{\mathrm {adS}}^{F,{\mathrm {RKT}}}(\beta ) - 3P_{\mathrm {adS}}^{F,{\mathrm {RKT}}}(\beta ) & = & \frac{2m^{3}\cos \rho }{\pi ^{2}\beta } \sum _{j=1}^{\infty } \frac{\left(-1\right) ^{j-1}}{j} K_{1}\left(\frac{mj\beta }{\cos \rho }\right)  ,
	\nonumber \\
	P_{\mathrm {adS}}^{F,{\mathrm {RKT}}}(\beta ) & = & \frac{2m^{2}\cos ^{2}\rho}{\pi ^{2}\beta ^{2}} \sum _{j=1}^{\infty } \frac{\left(-1\right) ^{j-1}}{j^{2}} K_{2}\left(\frac{mj\beta }{\cos \rho }\right) .
	\label{eq:fermionRKT}
	\end{eqnarray}	
\end{subequations}
In the massless limit $m\rightarrow 0$, instead of (\ref{eq:massless_M}), we now have
\begin{eqnarray}
\left. E_{\mathrm {adS}}^{S,{\mathrm {RKT}}}(\beta )\right\rfloor _{m=0} &  = &  \left. 3P_{\mathrm {adS}}^{S,{\mathrm {RKT}}}(\beta )\right\rfloor _{m=0} =\frac{\pi ^{2}\cos ^{4}\rho }{30\beta^4} ,
\nonumber \\
\left. E_{\mathrm {adS}}^{F,{\mathrm {RKT}}}(\beta )\right\rfloor _{m=0}  & = &  \left. 3P_{\mathrm {adS}}^{F,{\mathrm {RKT}}}(\beta )\right\rfloor _{m=0} =\frac{7\pi ^{2}\cos ^{4}\rho }{60 \beta^4} .
\label{eq:masslessRKT}
\end{eqnarray}
It is clear from (\ref{eq:adSRKT}, \ref{eq:masslessRKT}) that the energy density $E_{\mathrm {adS}}^{S}(\beta )$ and pressure $P_{\mathrm {adS}}^{S}(\beta )$, as anticipated, depend on the radial coordinate $\rho $ since the local inverse temperature ${\tilde {\beta }}$ (\ref{eq:tolman}) also depends on $\rho $.
In particular, both the energy density and pressure vanish as $\rho \rightarrow \pi /2$, and the boundary is approached.
At the origin $\rho =0$, the expressions (\ref{eq:adSRKT}, \ref{eq:masslessRKT}) reduce to those in Minkowski space-time (\ref{eq:RKT_M}, \ref{eq:massless_M}). 
The energy density and pressure also depend on the particle mass $m$, but not on the adS radius of curvature $a$. 

\begin{figure}
	\begin{center}
	\includegraphics[width=6cm]{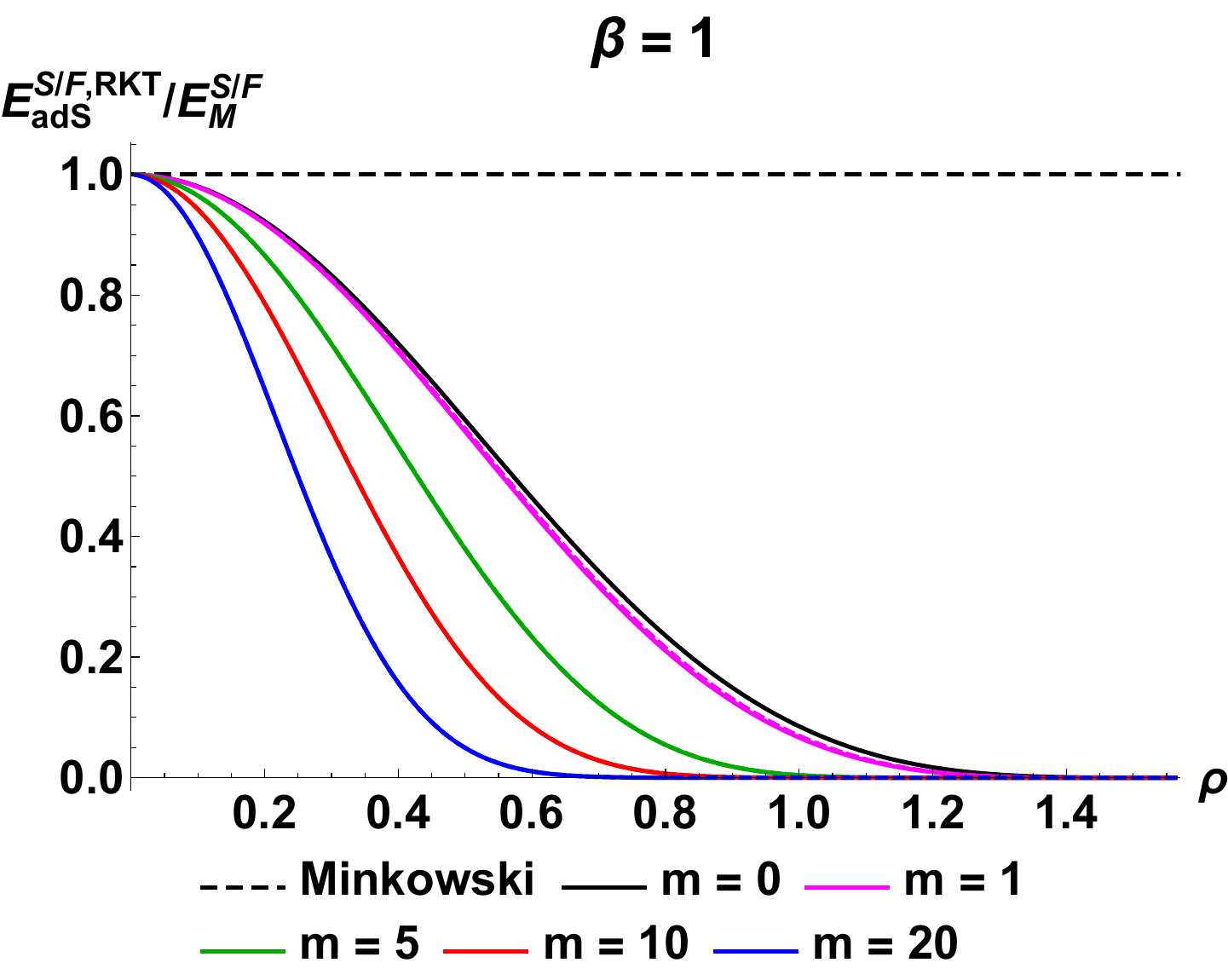}
	\includegraphics[width=6cm]{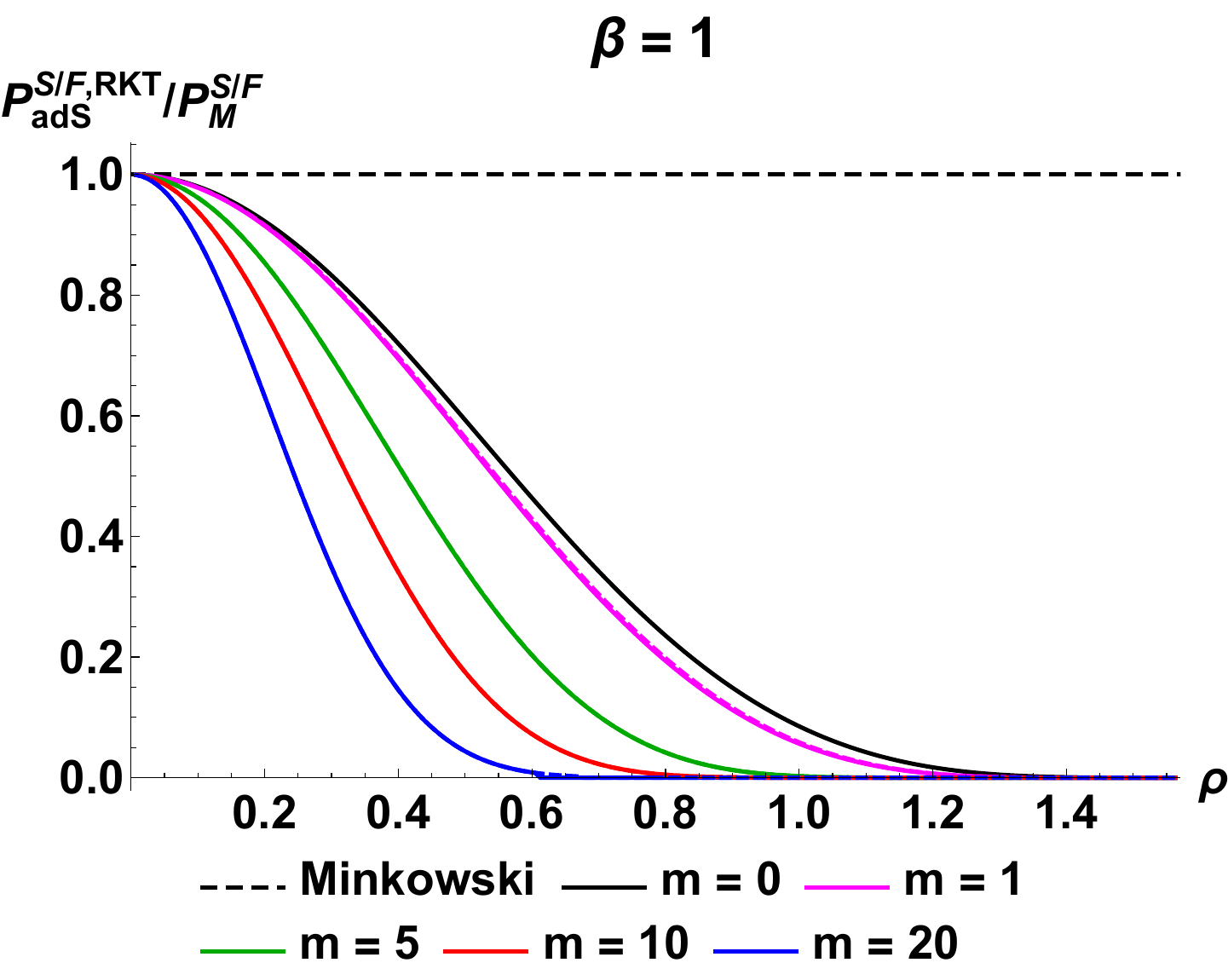}
	\end{center}
	\caption{Energy density $E_{\mathrm {adS}}^{S/F,{\mathrm {RKT}}}$ (left) and pressure  $P_{\mathrm {adS}}^{S/F,{\mathrm {RKT}}}$ (right) for a thermal gas of particles at inverse temperature $\beta =1$, divided by the corresponding Minkowski space-time values (\ref{eq:RKT_M}). Results are given for a selection of values of the particle mass $m$.  Solid curves denote scalar quantities and dashed curves those for fermions. \label{fig:RKT}}
\end{figure}

In Fig.~\ref{fig:RKT} we compare the behaviour of the energy density (left) and pressure (right) for scalars (solid lines) and fermions (dashed lines)  on adS with those on Minkowski space (\ref{eq:RKT_M}), in each case dividing the energy/pressure values on adS by the corresponding values on Minkowski space-time (\ref{eq:RKT_M}). For simplicity, dimensionless analogues of the dimensionful quantities $\beta $, $m$, $E$ and $P$ are introduced by multiplying each of these by a suitable power of the adS radius of curvature $a$. We then set $a=1$. We also fix the inverse temperature $\beta =1$ and consider a selection of values of the particle mass $m$.  

At the origin, the adS values are the same as those on Minkowski space-time, but, as $\rho $ increases, both the energy density and pressure decrease, as expected, vanishing on the space-time boundary at $\rho =\pi /2$ where the local inverse temperature (\ref{eq:tolman}) is zero.
As the particle mass $m$ increases, the profiles decrease more rapidly to zero as $\rho $ increases. In Fig.~\ref{fig:RKT} we have divided the energy density and pressure values in adS by those in Minkowski space, with the result that the curves for scalars and fermions are barely distinguishable. We find similar results for other values of the inverse temperature $\beta $. From this, we deduce that the difference between the energy densities and pressures in RKT for  Bose-Einstein and Fermi-Dirac statistics is not significantly influenced by the presence of curvature in adS.

In the next two sections we study the QFT of scalar and fermion fields on adS, and compare the quantum t.e.v.s with the RKT results presented in this section.

\section{Quantum scalar field on adS space-time}
\label{sec:scalar}

We first consider a quantum scalar field $\Phi $ of mass $m$, satisfying the Klein-Gordon equation
\begin{equation}
\left[ \Box -m^{2}-\xi R  \right] \Phi = 0 
\end{equation}
where $R$ is the Ricci scalar  (\ref{eq:Ricci}) and the coupling constant $\xi $ takes the value $\xi =1/6$ when the field is conformally coupled to the space-time curvature, and $\xi=0$ for minimal coupling. 
To simplify the algebra, it is helpful to introduce the quantity \cite{Kent:2014nya}
\begin{equation}
\eta = {\sqrt {m^{2}a^{2}+ \xi R a^{2} + \frac{9}{4}}},
\end{equation}
which takes the values $\eta = 1/2$ when the field is massless and conformally coupled, and $\eta = 3/2$ when the field is massless and minimally coupled.

\subsection{Vacuum expectation values}
\label{sec:scalarvac}

Our starting point for finding v.e.v.s is the vacuum Feynman Green's function, which satisfies the inhomogeneous Klein-Gordon equation
\begin{equation}
\left[ \Box _{x} -m^{2}-\xi R  \right] 
G_{{\mathrm {vac}}}^{\mathrm {adS}}(x,x') =\left( -g \right)^{-\frac{1}{2}} \delta ^{4}(x-x') ,
\label{eq:KGinhomo}
\end{equation}
where $g$ is the determinant of the metric (\ref{eq:adSmetric}).
The maximal symmetry of adS enables us to write the vacuum Feynman Green's function in terms of the geodetic interval $s_{\rm {adS}}(x,x')$ between two space-time points $x=(t,{\bm {x}})$ and $x'=(t',{\bm {x'}})$, which for adS takes the form
\begin{equation}
\cos \left(\frac {s_{\mathrm {adS}}}{a}\right) = \frac{\cos \Delta \tau }{\cos \rho \cos \rho '} - \cos {\bm {\gamma }} \tan \rho \tan \rho ',
\label{eq:sadS}
\end{equation}
where $\Delta \tau = \tau - \tau'$ and ${\bm  {\gamma }}$ is the angular separation of the points, given by
\begin{equation}
\cos {\bm {\gamma }} = \frac{{\bm {x \cdot x'}}}{\rho \rho'}.
\end{equation} 
Following \cite{Allen:1985wd,Camporesi:1991nw}, the vacuum Feynman Green's function can be written in terms of $s_{\mathrm {adS}}$ in the form
\begin{eqnarray}
-iG_{\mathrm {vac}}^{\mathrm {adS}}(x,x') & = &  {\mathcal {C}}_{\mathrm {S}} \left[ -\sin \left(\frac {s_{\mathrm {adS}}}{2a} \right) \right]^{-3-2\eta } {}_{2}F_{1} \left[ \frac{3}{2}+ \eta , \frac{1}{2}+\eta ; 1+2\eta ; {\mathrm {cosec}}^{2} \left(\frac{s_{\mathrm {adS}}}{2a} \right) \right] \nonumber \\
& &
\hspace{-0.5cm}
+{\mathcal {D}}_{\mathrm {S}}\left[ -\sin \left(\frac {s_{\mathrm {adS}}}{2a} \right) \right]^{-3+2\eta } {}_{2}F_{1} \left[ \frac{3}{2}- \eta , \frac{1}{2}-\eta ; 1-2\eta ; {\mathrm {cosec}}^{2} \left(\frac{s_{\mathrm {adS}}}{2a} \right) \right] ,
\nonumber \\
\label{eq:GvacSadS}
\end{eqnarray}
where ${\mathcal {C}}_{\mathrm {S}}$ and ${\mathcal {D}}_{\mathrm {S}}$ are arbitrary constants and ${}_{2}F_{1}[*,*;*;*]$ is a hypergeometric function. 
The requirement that the vacuum Feynman Green's function be regular when $\rho '\rightarrow \pi /2$ and the point $x'$ approaches the adS boundary implies that the constant ${\mathcal {D}}_{\mathrm {S}}=0$.
This condition corresponds to Dirichlet boundary conditions on the scalar field at the space-time boundary, or, alternatively, considering only ``regular'' modes of the scalar field \cite{Camporesi:1991nw}, and can be applied to the Green's function for any value of $\eta $. For some values of $\eta $, it is possible to also consider Neumann boundary conditions at the adS boundary \cite{Avis:1977yn}.
The constant ${\mathcal {C}}_{\mathrm {S}}$ is fixed by requiring that the leading-order short-distance singularity structure of the vacuum Feynman Green's function (\ref{eq:GvacSadS}) matches that for the scalar vacuum Feynman Green's function on Minkowski space-time. This gives \cite{Kent:2014nya}
\begin{equation}
{\mathcal {C}}_{\mathrm {S}} = 
-\frac{1}{8\pi a^{2}}\frac{\Gamma (-2\eta )}{\Gamma \left( \frac{3}{2}- \eta \right)\Gamma \left(\frac{1}{2}-\eta \right)}\left(\frac{1}{4}-\eta ^{2} \right)\tan \left(\pi \eta \right) =
\begin{cases}
\frac{1}{16\pi ^{2} a^{2}}, & {\mbox {for }} \eta = \frac{1}{2}, \\
\frac{1}{48\pi ^{2}a^{2}}, & {\mbox {for }} \eta = \frac{3}{2},
\end{cases}
\end{equation}
where $\Gamma (*)$ is the usual Gamma function.

To find the v.e.v., we apply the curved space-time analogue of the Minkowski stress energy tensor operator (\ref{eq:scalar_set_M}) to the vacuum Feynman Green's function (\ref{eq:GvacSadS}) and bring the space-time points together:
\begin{eqnarray}
\braket{T_{{\hat {\alpha }}{\hat {\nu}}}}^{{\mathrm {S}},\mathrm {adS}}_{\mathrm {vac}}
& = & \lim_{x'\rightarrow x}  \left\{ \left[ 
\left( 1- 2\xi \right)  g_{{\hat {\nu }}}{}^{{\hat {\nu }}'} \nabla _{{\hat {\alpha }}} \nabla_{{\hat {\nu }}'}
+ \left(2\xi - \frac{1}{2}\right) g_{{\hat {\alpha }}{\hat {\nu }}} g^{{\hat {\lambda }} {\hat { \sigma }}'} \nabla _{{\hat {\lambda }}}\nabla _{\hat {\sigma }'}
\right. \right.
\nonumber \\ & & 
\left. \left. \hspace{-1cm}
-2\xi g_{{\hat {\alpha }}}{}^{{\hat {\alpha }}'} g_{{\hat {\nu}}}{}^{{\hat {\nu }}'} \nabla _{{\hat {\alpha }}'}\nabla _{{\hat {\nu }}'} 
+2\xi g_{{\hat {\alpha }}{\hat {\nu }}} \Box _{x}
+ \xi \left( R_{{\hat {\alpha }}{\hat {\nu }}} - \frac{1}{2}g_{{\hat {\alpha }}{\hat {\nu }}}R\right) -\frac{1}{2}g_{{\hat {\alpha }}{\hat {\nu }}}m^{2} 
\right] iG^{\mathrm {adS}}_{\rm {vac}}
\right\} ,
\nonumber \\
\label{eq:SET_scalar}
\end{eqnarray}
where $g_{{\hat {\alpha }}}{}^{{\hat {\nu }}'}$ is the bivector of parallel transport (whose components on adS space-time can be found explicitly in \cite{Ambrus:2017cow}). 
As on Minkowski space-time, this procedure yields an infinite quantity. Unlike the situation on Minkowski space-time, we cannot simply set this quantity to zero.
This is because the expectation value of the SET appears on the right-hand side of the semi-classical Einstein equations (\ref{eq:SCEE}), and hence its value has to be obtained in a consistent manner.  An appropriate method of renormalization therefore has to be applied to the computation of v.e.v.s.

In this report we adopt Hadamard renormalization as our method of obtaining finite v.e.v.s.  
In this approach, the short-distance singularity structure of the Feynman Green's function is given by the Hadamard parametrix (states for which this is the case are referred to as Hadamard states).
For a quantum scalar field in four space-time dimensions, the Hadamard parametrix for the singular part of the Feynman Green's function is \cite{Decanini:2005eg}
\begin{equation}	
-iG_{{\rm {Had}}}(x,x')= \frac{1}{8\pi ^{2}} \left[ \frac{\Delta (x,x')^{\frac{1}{2}}}{\sigma } +V(x,x') \ln \left(  M^{2}\sigma \right) \right] ,
\label{eq:scalarH}
\end{equation} 
where it is understood that $\sigma $ should be replaced by $\sigma +i\epsilon $ for $\epsilon >0$ arbitrarily small and $M$ is a mass renormalization scale.
Here the world function $\sigma $ is related to the geodetic interval $s_{\mathrm {adS}}$ (\ref{eq:sadS}) by
\begin{equation}
2\sigma = -s_{\mathrm {adS}}^{2},
\end{equation}
and $\Delta (x,x')$ is the van-Vleck-Morette determinant, which on four-dimensional adS has the closed form \cite{Kent:2014nya}
\begin{equation}
\Delta (x,x') = \left(\frac{s_{\mathrm {adS}}}{a}\right) ^{3} {\mathrm {cosec}} ^{3} \left( \frac{s_{\mathrm {adS}}}{a}\right) .
\label{eq:VVMD}
\end{equation}
The quantity $V(x,x')$ is a biscalar which is regular as $x\rightarrow x'$ and satisfies the homogeneous Klein-Gordon equation.
Due to the maximal symmetry of adS, it must be the case that $V(x,x')$ depends only on the geodetic interval $s_{\mathrm {adS}}$ (\ref{eq:sadS}). We write the solution of the homogeneous Klein-Gordon equation in the form
\begin{eqnarray}
V(x,x') & = &   {\mathcal {C}}_{V} {}_{2}F_{1} \left[ \frac{3}{2}+\eta , \frac{3}{2}-\eta ; 2; \sin ^{2} \left(\frac{s_{\mathrm {adS}}}{a} \right) \right] 
\nonumber \\ & & 
+{\mathcal {D}}_{V} {}_{2}F_{1} \left[  \frac{3}{2}+\eta , \frac{3}{2}-\eta ; 2; \cos ^{2} \left(\frac{s_{\mathrm {adS}}}{a} \right)\right] 
\end{eqnarray}
where ${\mathcal {C}}_{V}$ and ${\mathcal {D}}_{V}$ are arbitrary constants.
The second term in $V(x,x')$ diverges as $s_{\mathrm {adS}}\rightarrow 0$ and therefore we set ${\mathcal {D}}_{V}=0$.
The constant ${\mathcal {C}}_{V}$ is fixed by applying the boundary conditions on $V(x,x')$ \cite{Decanini:2005eg}, which give \cite{Kent:2014nya}
\begin{equation}
{\mathcal {C}}_{V} = \lim _{s_{\mathrm {adS}}\rightarrow 0} \left[ \frac{1}{2}\left( m^{2}+\xi R - \Box _{x}\right)  \Delta (x,x')^{\frac{1}{2}}\right] = \frac{1}{8a^{2}} \left(4\eta ^{2}-1 \right) .
\label{eq:CVBC}
\end{equation}
For a massless, conformally coupled scalar field with $\eta = \frac{1}{2}$, we see that $V(x,x')$ is identically zero, while, for a massless minimally coupled scalar field with $\eta = \frac{3}{2}$, we have $V(x,x')\equiv a^{-2}$, which is a constant. 

Whatever the scalar field mass and coupling, the Hadamard parametrix (\ref{eq:scalarH}) is purely geometric and independent of the quantum state under consideration.
In the Hadamard renormalization prescription, the Hadamard parametrix (\ref{eq:scalarH}) is subtracted from the Feynman Green's function (\ref{eq:GvacSadS}) before the stress-energy tensor operator (\ref{eq:SET_scalar}) is applied and the points brought together.  This gives us the renormalized v.e.v.
\begin{eqnarray}
\braket{T_{{\hat {\alpha }} {\hat {\nu }}}}^{{\mathrm {S}},\mathrm {adS}}_{\mathrm {vac}}
& = & \lim_{x'\rightarrow x}  \left\{ \left[ 
\left( 1- 2\xi \right)  g_{{\hat {\nu}}}{}^{{\hat {\nu}}'} \nabla _{{\hat {\alpha }}} \nabla_{{\hat {\nu }}'}
+ \left(2\xi - \frac{1}{2}\right) g_{{\hat {\alpha }}{\hat {\nu}}} g^{{\hat {\lambda }} {\hat \sigma }'} \nabla _{{\hat {\lambda }}}\nabla _{\hat {\sigma }'}
\right. \right.
\nonumber \\ & & 
\left. \left. 
-2\xi g_{{\hat {\alpha }}}{}^{{\hat {\alpha }}'} g_{{\hat {\nu }}}{}^{{\hat {\nu}}'} \nabla _{{\hat {\alpha }}'}\nabla _{{\hat {\nu }}'} 
+2\xi g_{{\hat {\alpha }}{\hat {\nu }}} \Box _{x} 
\right. \right. \nonumber \\ & & \left.\left. 
+ \xi \left( R_{{\hat {\alpha }}{\hat {\nu }}} - \frac{1}{2}g_{{\hat {\alpha }}{\hat {\nu }}}R\right) -\frac{1}{2}g_{{\hat {\alpha }}{\hat {\nu }}}m^{2} 
\right] \left[ iG^{\mathrm {adS}}_{\rm {vac}} - iG_{\mathrm {Had}} \right] 
\right\} .
\nonumber \\
\label{eq:scalar_SET}
\end{eqnarray}
An explicit computation (the details of which can be found in \cite{Kent:2014nya}) gives the v.e.v.~of the scalar field SET in closed form:
\begin{eqnarray}
\braket{T_{\hat{\alpha}}^{\hat{\nu}}}^{{\mathrm {S}},\mathrm {adS}}_{\mathrm {vac}} & = &
\left\{ \frac{3}{128 \pi ^{2}a^{4}} \left[-\frac{4}{3}\eta ^{4}  - \left( 16\xi - \frac{10}{3} \right) \eta ^{2} +4\xi - \frac{3}{4} \right] \Upsilon _{S} 
\right. \nonumber \\ & &  \left.
+ \frac{3}{128 \pi ^{2}a^{4}}\left[ \eta ^{4} + \left(8\xi - \frac{29}{18}\right) \eta ^{2} 
+ \frac{2}{3}\xi -\frac {107}{720}\right] 
\right. \nonumber \\ & &  \left.
+ \frac{\ln M^{2}}{64\pi ^{2}a^{4}}\left[\eta ^{4} +\left(12\xi - \frac{5}{2}\right)\eta ^{2} -3\xi +\frac{9}{16}\right]
\right\} \delta _{\hat {\alpha }}^{\hat {\nu }} ,
\label{eq:vevSadS}
\end{eqnarray}
where
\begin{equation}
		\Upsilon _{S}=   \psi \left( \frac{1}{2}+ \eta \right) + C  -\ln \left(2Ma \right)  , 
\end{equation}
with $C$ the Euler-Mascheroni constant and $\psi (*) = \Gamma '(*)/\Gamma (*)$ is the digamma function.
The result (\ref{eq:vevSadS}) agrees with that obtained using $\zeta$-function regularization \cite{Camporesi:1992wn}.

For all values of the scalar field mass $m$ and coupling $\xi$, the tetrad components of the v.e.v.~of the SET (\ref{eq:vevSadS}) are constants (so that the v.e.v.~of the SET in coordinate components is proportional to the metric tensor), which is expected since adS space-time is maximally symmetric.
In general, the v.e.v.~of the SET depends on the unknown mass renormalization scale $M$. However, for a massless scalar field either conformally or minimally coupled to the space-time curvature, the coefficient of $\ln M$ in (\ref{eq:vevSadS}) vanishes and the v.e.v.~simplifies to
\begin{subequations}
	\label{eq:scalar_vevs}
\begin{equation}
\braket{T_{\hat{\alpha}}^{\hat{\nu}}}^{{\mathrm {S}},\mathrm {adS}}_{\mathrm {vac,cc}} = 
- \frac{1}{960 \pi ^{2}a^{4}}\delta _{\hat {\alpha }}^{\hat {\nu }} 
\end{equation}
for conformal coupling, and 
\begin{equation}
\braket{T_{\hat{\alpha}}^{\hat{\nu}}}^{{\mathrm {S}},\mathrm {adS}}_{\mathrm {vac,mc}} =
\frac{29}{960 \pi ^{2}a^{4}}\delta _{\hat {\alpha }}^{\hat {\nu}} 
\end{equation}
\end{subequations}
for minimal coupling.

\subsection{Thermal expectation values}
\label{sec:scalarT}

Having found the v.e.v.~of the SET for a quantum scalar field on adS, we now turn our attention to the calculation of the t.e.v.~at inverse temperature $\beta $.
For a massless, conformally coupled, scalar field on adS, the t.e.v.~of the SET was computed in \cite{Allen:1986ty} using the fact that the thermal Feynman Green's function in this case is a meromorphic function which is doubly periodic in time $\tau $ (one period results from the periodicity of the coordinate $\tau $ on adS, see Sec.~\ref{sec:adSgeometry}, and the second from the periodicity in imaginary time for a thermal state, see (\ref{eq:scalar_thermal_ads_G}) below).
This double-periodicity enables the thermal Feynman Green's function to be written in terms of an elliptic function, which simplifies the calculation \cite{Allen:1986ty}.
However, this method can only be applied to a massless, conformally coupled, scalar field for which the biscalar $V(x,x')$ in the Hadamard parametrix (\ref{eq:scalarH}) vanishes identically.  If $V(x,x')$ is nonzero, then the vacuum Feynman Green's function (and hence the thermal Feynman Green's function) possesses a branch cut, and is no longer meromorphic.
This implies that in general the thermal Feynman Green's function cannot be written in terms of an elliptic function.

In this section we take an alternative approach, computing the t.e.v.~of the SET directly from the thermal Feynman Green's function without using elliptic functions. Our approach is valid for all values of the scalar field mass $m$ and the coupling constant $\xi $.  When $m=0$ and $\xi = 1/6$, we obtain the same numerical values for the t.e.v.~of the SET as those in \cite{Allen:1986ty}.  
The thermal scalar Feynman Green's function $G_{\beta }^{\mathrm {adS}}(x,x')$ is constructed from the vacuum Feynman Green's function $G_{\mathrm {vac}}^{\mathrm {adS}}(x,x')$ in the same way as in Minkowski space-time (\ref{eq:scalar_thermal_M}):
\begin{equation}
G_{\beta }^{\rm {adS}}(x,x') =  
\sum _{j=-\infty }^{\infty } G_{\rm {vac}}^{\rm {adS}} (\tau +ij{\bar {\beta }},{\bm {x}}; \tau ',\bm {x'}) ,
\label{eq:scalar_thermal_ads_G}
\end{equation}
where 
\begin{equation}
{\bar {\beta}} =\beta a^{-1}
\label{eq:betabar}
\end{equation} is a dimensionless quantity, introduced since 
our temporal coordinate $\tau $ is also dimensionless. 
To find the t.e.v.~of the SET, we would need to apply the SET operator (\ref{eq:SET_scalar}) to the thermal Feynman Green's function (\ref{eq:scalar_thermal_ads_G}) and bring the space-time points together.  As usual, this gives an infinite answer which requires renormalization. We saw in Sec.~\ref{sec:scalarvac} that the Hadamard parametrix (\ref{eq:scalarH}) used to renormalize the v.e.v.~of the SET is purely geometric and does not depend on the quantum state under consideration. Therefore the t.e.v.~is renormalized by subtracting the same Hadamard parametrix from the thermal Feynman Green's function.
As a result, the difference between the t.e.v.~and the v.e.v. of the SET, given by 
\begin{eqnarray}
\braket{:T_{\hat{\alpha}\hat{\nu}}:}_{\beta }^{\mathrm {S,adS}}  & \equiv &
\braket{T_{\hat{\alpha}\hat{\nu}}}_{\beta }^{\mathrm {S,adS}} -
\braket{T_{\hat{\alpha}\hat{\nu}} }_{\rm {vac}}^{\mathrm {S,adS}} 
\nonumber \\ & =  & 
\lim_{x'\rightarrow x}  \left\{ \left[ 
\left( 1- 2\xi \right)  g_{{\hat {\nu }}}{}^{{\hat {\nu}}'} \nabla _{{\hat {\alpha }}} \nabla_{{\hat {\nu }}'}
+ \left(2\xi - \frac{1}{2}\right) g_{{\hat {\alpha }}{\hat {\nu }}} g^{{\hat {\lambda }} {\hat \sigma }'} \nabla _{{\hat {\lambda }}}\nabla _{\hat {\sigma }'}
\right. \right.
\nonumber \\ & & 
\left. \left. 
-2\xi g_{{\hat {\alpha }}}{}^{{\hat {\alpha }}'} g_{{\hat {\nu}}}{}^{{\hat {\nu}}'} \nabla _{{\hat {\alpha }}'}\nabla _{{\hat {\nu }}'} 
+2\xi g_{{\hat {\alpha }}{\hat {\nu }}} \Box _{x}
\right.\right. \nonumber \\ & & \left.\left.
+ \xi \left( R_{{\hat {\alpha }}{\hat {\nu }}} - \frac{1}{2}g_{{\hat {\alpha }}{\hat {\nu }}}R\right) -\frac{1}{2}g_{{\hat {\alpha }}{\hat {\nu }}}m^{2} 
\right] \left[ iG^{\mathrm {adS}}_{\beta } - iG^{\mathrm {adS}}_{\mathrm {vac}} \right] 
\right\} ,
\label{eq:scalar_thermal_adS}
\end{eqnarray}
does not require renormalization.  We therefore consider this difference, which will facilitate comparison with both the Minkowski space-time results of Sec.~\ref{sec:Minkowski} and the RKT results from Sec.~\ref{sec:RKT}.
In this section we restrict our attention to massless scalar fields with either minimal or conformal coupling to the space-time curvature.  Results for massive fields and more general curvature couplings will be presented in a forthcoming paper \cite{KentEW}.

As on Minkowski space-time, subtracting the vacuum Feynman Green's function from the thermal Feynman Green's function simply means removing the $j=0$ term from the sum in (\ref{eq:scalar_thermal_ads_G}). 
In general the algebraic expressions for the components of the difference between the t.e.v.~and v.e.v.~of the SET computed from (\ref{eq:scalar_thermal_adS}) are extremely lengthy. Here we therefore present the results for a massless scalar field either conformally or minimally coupled to the Ricci scalar curvature.
When the scalar field is massless and conformally coupled,  we find the following components of (\ref{eq:scalar_thermal_adS}):
\begin{subequations}
	\label{eq:SET_thermal_cc}
\begin{eqnarray}
\braket{:T^{{\hat {\tau }}{\hat {\tau }}}:}_{\beta ,{\mathrm {cc}}}^{\mathrm {S,adS}}
& = & 
\frac{\cos ^{4}\rho }{192 \pi ^2 a^{4} }\sum _{j=1}^{\infty }
\frac{{\mathcal {F}}^{\rm {cc}}_{\tau }}{\left[\cos(2 \rho )+\cosh(j {\bar {\beta}})\right] ^3 \sinh ^{4}\left( \frac{j{\bar {\beta}}}{2}\right)}  ,
\label{eq:E_thermal_cc}
\\
\braket{:T^{{\hat {\rho}}{\hat {\rho }}}:}_{\beta ,{\mathrm {cc}}}^{\mathrm {S,adS}}
& = &
\frac{\cos ^{4}\rho }{96 \pi ^2 a^{4} }\sum _{j=1}^{\infty }
\frac{{\mathcal {F}}^{\rm {cc}}_{\rho }}{\left[ \cos(2 \rho )+\cosh(j {\bar {\beta}})\right] ^2 \sinh ^{4}\left( \frac{j{\bar {\beta}}}{2}\right)},
\\
 \braket{:T^{\hat {\theta }{\hat {\theta  }}}:}_{\beta ,{\mathrm {cc}}}^{\mathrm {S,adS}}
 & = & 
 \frac{\cos ^{4}\rho }{192 \pi ^2 a^{4} }\sum _{j=1}^{\infty }
 \frac{{\mathcal {F}}^{\rm {cc}}_{\theta  }}{\left[ \cos(2 \rho )+\cosh (j {\bar {\beta}}) \right] ^3 \sinh ^{4}\left( \frac{j{\bar {\beta}}}{2}\right)},
\end{eqnarray}
where 
\begin{eqnarray}
{\mathcal {F}}_{\tau }^{\rm {cc}} & = &  3 \cos(6 \rho ) \left[ 2+\cosh(j {\bar {\beta}})\right] +18 \cos(4 \rho ) \cosh(j{\bar {\beta}}) \left[2+\cosh(j {\bar {\beta}})\right] 
\nonumber \\ & & 
+\cos(2 \rho ) \left[ 44+51 \cosh(j {\bar {\beta}})+30 \cosh (2j {\bar {\beta}})+10 \cosh(3 j {\bar {\beta}}) \right]
\nonumber \\ & & 
+ 31+33 \cosh(j {\bar {\beta}}) +15 \cosh(2 j {\bar {\beta}})+11 \cosh(3 j {\bar {\beta}}) , 
\\
{\mathcal {F}}_{\rho }^{\rm {cc}} & = &  \cos(4 \rho )\left[2 + \cosh (j{\bar {\beta}}) \right]
+ 4 \cos (2\rho )\cosh (j{\bar {\beta}})\left[ 2+\cosh(j {\bar {\beta}})\right] 
\nonumber \\ & & 
+ 9 -5\cosh(j {\bar {\beta}}) +5 \cosh( 2 j {\bar {\beta}}) ,
\\
{\mathcal {F}}^{\rm {cc}}_{\theta  } & = & 
\cos(6 \rho ) \left[ 2+\cosh( j {\bar {\beta}}) \right]
+6 \cos(4 \rho ) \cosh(j {\bar {\beta}}) \left[ 2+\cosh(j {\bar {\beta}})\right] 
\nonumber \\ & & 
+ \cos(2 \rho ) \left[ 8+27 \cosh(j {\bar {\beta}})+6 \cosh(2 j {\bar {\beta}})+4 \cosh (3 j {\bar {\beta}}
) \right]  
\nonumber \\  & & 
+17+ \cosh(j {\bar {\beta}})
+9 \cosh(2 j {\bar {\beta}})+3 \cosh (3 j {\bar {\beta}}),
\end{eqnarray}
\end{subequations}
and $\braket{:T^{{\hat {\varphi }}{\hat {\varphi }}}:}_{\beta ,{\mathrm {cc}}}^{\mathrm {S,adS}}=\braket{:T^{\hat {\theta }{\hat {\theta  }}}:}_{\beta ,{\mathrm {cc}}}^{\mathrm {S,adS}}$.
In the massless, minimally coupled case, the components of (\ref{eq:scalar_thermal_adS}) are algebraically simpler:
\begin{subequations}
	\label{eq:SET_thermal_mc}
	\begin{eqnarray}
	\braket{:T^{\hat {\tau }{\hat {\tau }}}:}_{\beta ,{\mathrm {mc}}}^{\mathrm {S,adS}} & = & 
	\frac{3 \cos ^{6}\rho }{8 \pi ^2 a^{4}} \sum _{j=1}^{\infty }
	\frac{1}{\left[ \cos(2 \rho )+\cosh(j {\bar {\beta}})\right] \sinh ^{4}\left( \frac{j{\bar {\beta}} }{2}\right)},
	\label{eq:E_thermal_mc}
	\\ 
	\braket{:T^{\hat {\rho }{\hat {\rho }}}:}_{\beta ,{\mathrm {mc}}}^{\mathrm {S,adS}} & = & 
	\frac{ \cos ^{6}\rho }{8 \pi ^2 a^{4}} \sum _{j=1}^{\infty }
	\frac{ \left[ -2+\cos(2 \rho )+3 \cosh (j {\bar {\beta}}) \right] }{\left[ \cos(2 \rho )+\cosh(j {\bar {\beta}})\right] ^2 \sinh ^{4}\left( \frac{j{\bar {\beta}}}{2}\right)},
	\\
	\braket{:T^{{\hat {\theta  }}{\hat {\theta }}}:}_{\beta ,{\mathrm {mc}}}^{\mathrm {S,adS}}
	& = & 
	\frac{\cos ^{6}\rho }{16 \pi ^2 a^{4}} \sum _{j=1}^{\infty }
	\frac{{\mathcal {F}}_{\theta  }^{\rm {mc}} }{ \left[ \cos(2 \rho )+\cosh(j {\bar {\beta}})\right]^3 \sinh ^{4}\left( \frac{j{\bar {\beta}}}{2}\right)} ,
	\end{eqnarray}
where 
\begin{equation}
{\mathcal {F}}_{\theta  }^{\rm {mc}}  = 
\cos (4\rho) \cosh (j{\bar {\beta}})  
 + \cos (2\rho ) \left[ 5 -4 \cosh (j{\bar {\beta}}) + 3\cosh (2j{\bar {\beta}}) \right] 
-4+ 7 \cosh(j{\bar {\beta}}) ,
\end{equation}
\end{subequations}
and again we have $\braket{:T^{\hat {\varphi }{\hat {\varphi }}}:}_{\beta ,{\mathrm {mc}}}^{\mathrm {S,adS}}=\braket{:T^{\hat {\theta }{\hat {\theta  }}}:}_{\beta ,{\mathrm {mc}}}^{\mathrm {S,adS}}$.

It is clear from (\ref{eq:SET_thermal_cc}, \ref{eq:SET_thermal_mc}) that, unlike the situation both on Minkowski space-time and in RKT on adS space-time, the spatial components of the t.e.v.~of the SET are not identical, as $\braket{:T^{\hat {\rho }{\hat {\rho }}}:}_{\beta }^{\mathrm {S,adS}}\neq \braket{:T^{\hat {\theta }{\hat {\theta  }}}:}_{\beta }^{\mathrm {S,adS}}$ for both conformal and minimal coupling.
In order to compare our QFT results derived in this section with those from RKT (and the corresponding results on Minkowski space-time), it is convenient to perform a Landau decomposition of the t.e.v.~of the SET \cite{Landau}.
In the Landau decomposition, we write the components of the t.e.v.~of the SET in the form
\begin{eqnarray}
\braket{:T^{\hat {\alpha }{\hat {\nu }}}:}_{\beta }^{\mathrm {S,adS}} & = &
{\rm {Diag}} \left\{  E_{\mathrm{adS}}^{S,{\mathrm {QFT}}} (\beta ), P_{\mathrm{adS}}^{S,{\mathrm {QFT}}}(\beta )+\Pi_{\mathrm{adS}}^{S,{\mathrm {QFT}}} (\beta ) , 
\right. \nonumber \\ & & \left. \hspace{-0.5cm}
P_{\mathrm{adS}}^{S,{\mathrm {QFT}}} (\beta ) - \frac{1}{2} \Pi_{\mathrm{adS}}^{S,{\mathrm {QFT}}} (\beta ) , P_{\mathrm{adS}}^{S,{\mathrm {QFT}}}(\beta ) - \frac{1}{2}\Pi_{\mathrm{adS}}^{S,{\mathrm {QFT}}} (\beta ) 
\right\} ,
\label{eq:Landau}
\end{eqnarray}
where $E_{\mathrm{adS}}^{S,{\mathrm {QFT}}}(\beta )$ is the energy density, $P_{\mathrm{adS}}^{S,{\mathrm {QFT}}}(\beta )$ the pressure and $\Pi _{\mathrm{adS}}^{S,{\mathrm {QFT}}}(\beta )$ the shear stress or pressure deviator.
We note that the same decomposition arises with respect to the $\beta$ 
frame, which is the frame in which the macroscopic four-velocity is taken to be that of a relativistic thermometer in thermal equilibrium with the fluid \cite{Becattini:2014yxa}. Here, the Landau velocity $u_L^{\hat{\alpha}} = (1,0,0,0)^T$ is proportional to the time-translation Killing vector corresponding to the adS Hamiltonian. This adS Hamiltonian gives rise to the density operator generating
the thermal state considered in this section and hence the Landau frame and $\beta $ frame coincide.

We therefore have $E_{\mathrm{adS,cc}}^{S,{\mathrm {QFT}}}(\beta ) = \braket{:T^{\hat {\tau }{\hat {\tau }}}:}_{\beta ,{\mathrm {cc}}}^{\mathrm {S,adS}}$ and  $E_{\mathrm{adS,mc}}^{S,{\mathrm {QFT}}}(\beta ) = \braket{:T^{\hat {\tau }{\hat {\tau }}}:}_{\beta ,{\mathrm {mc}}}^{\mathrm {S,adS}}$,
while the expressions for the pressure and pressure deviator, in the conformally and minimally coupled cases, are, respectively,
\begin{subequations}
	\label{eq:QFTscalarP}
\begin{eqnarray}
P_{\mathrm{adS,cc}}^{S,{\mathrm {QFT}}}(\beta ) & = & 
\frac{\cos ^{4}\rho }{576 \pi ^2 a^{4} }\sum _{j=1}^{\infty }
\frac{{\mathcal {P}}_{\rm {cc}}}{\left[ \cos(2 \rho )+\cosh(j {\bar {\beta}})\right] ^3 \sinh ^{4}\left( \frac{j{\bar {\beta}}}{2}\right)},
\\
\Pi _{\mathrm{adS,cc}}^{S,{\mathrm {QFT}}}(\beta ) & = & 
\frac {4 \cos^4 \rho  \sin ^{2}\rho}{9\pi ^{2}a^{4}}
\sum _{j=1}^{\infty } \frac{\sinh ^{2}\left(\frac{j{\bar {\beta}}}{2}\right)}{\left[ \cos(2 \rho )+\cosh(j {\bar {\beta}})\right] ^3}  , 
\\
P_{\mathrm{adS,mc}}^{S,{\mathrm {QFT}}}(\beta ) & = & \frac{\cos ^{6}\rho }{24 \pi ^2 a^{4} }\sum _{j=1}^{\infty }
\frac{{\mathcal {P}}_{\rm {mc}}}{\left[ \cos(2 \rho )+\cosh(j {\bar {\beta}})\right] ^3 \sinh ^{4}\left( \frac{j{\bar {\beta}}}{2}\right)},
\\
\Pi _{\mathrm{adS,mc}}^{S,{\mathrm {QFT}}}(\beta ) & = & 
\frac { \cos^{6} \rho  \sin ^{2}\rho}{3\pi ^{2}a^{4}}
\sum _{j=1}^{\infty } \frac{
-2+\cos \left(2\rho \right) +3\cosh \left(j{\bar {\beta}}\right)}{\left[ \cos(2 \rho )+\cosh(j {\bar {\beta}})\right] ^3\sinh ^{2}\left(\frac{j{\bar {\beta}}}{2}\right)} , 
\end{eqnarray}
where
\begin{eqnarray}
{\mathcal {P}}_{\mathrm {cc}} & = &
3\cos \left( 6\rho \right) \left[2 + \cosh \left( j{\bar {\beta}}\right)  \right] 
+18 \cos \left( 4\rho \right)  \cosh \left( j{\bar {\beta}} \right)\left[2 + \cosh \left( j{\bar {\beta}} \right)  \right]  
\nonumber \\ & & 
+\cos (2 \rho) \left[ 44 + 51 \cosh\left( j {\bar {\beta}} \right) + 30 \cosh \left( 2 j {\bar {\beta}} \right) + 
10 \cosh \left( 3 j {\bar {\beta}} \right)  \right] 
\nonumber \\ & & 
+31+ 33 \cosh \left( j{\bar {\beta}}\right) + 15 \cosh \left( 2j{\bar {\beta}}\right) + 11\cosh \left( 3j{\bar {\beta}}\right) ,
\\
{\mathcal {P}}_{\mathrm {mc}} & = &
\frac{1}{2}\cos \left( 4\rho \right)\left[ 1 + 2 \cosh \left(j{\bar {\beta}}\right) \right] 
+ 3\cos \left( 2\rho \right) \left[ 1 + \cosh \left( 2j{\bar {\beta}}\right) \right]
\nonumber \\ & & 
-2 + 5 \cosh \left( j{\bar {\beta}}\right) + \frac{3}{2}\cosh \left( 2j{\bar {\beta}}\right) .
\end{eqnarray}
\end{subequations}
In both the conformally and minimally coupled cases, the pressure deviator $\Pi _{\mathrm{adS}}^{S,{\mathrm {QFT}}}(\beta )$ is nonzero. This is our first indication that the difference between the t.e.v.~and v.e.v.~in QFT is not well approximated by the results from RKT on adS space-time presented in Sec.~\ref{sec:RKT}.
This term cannot be present in classical states which are in thermal equilibrium and hence arises solely as a quantum correction to the classical (RKT) SET, as also remarked in 
\cite{Becattini:2015nva,Panerai:2015xlr,Ambrus:2017opa}.

\begin{figure}
	\begin{center}
		\begin{tabular}{cc}
		\includegraphics[width=6.5cm]{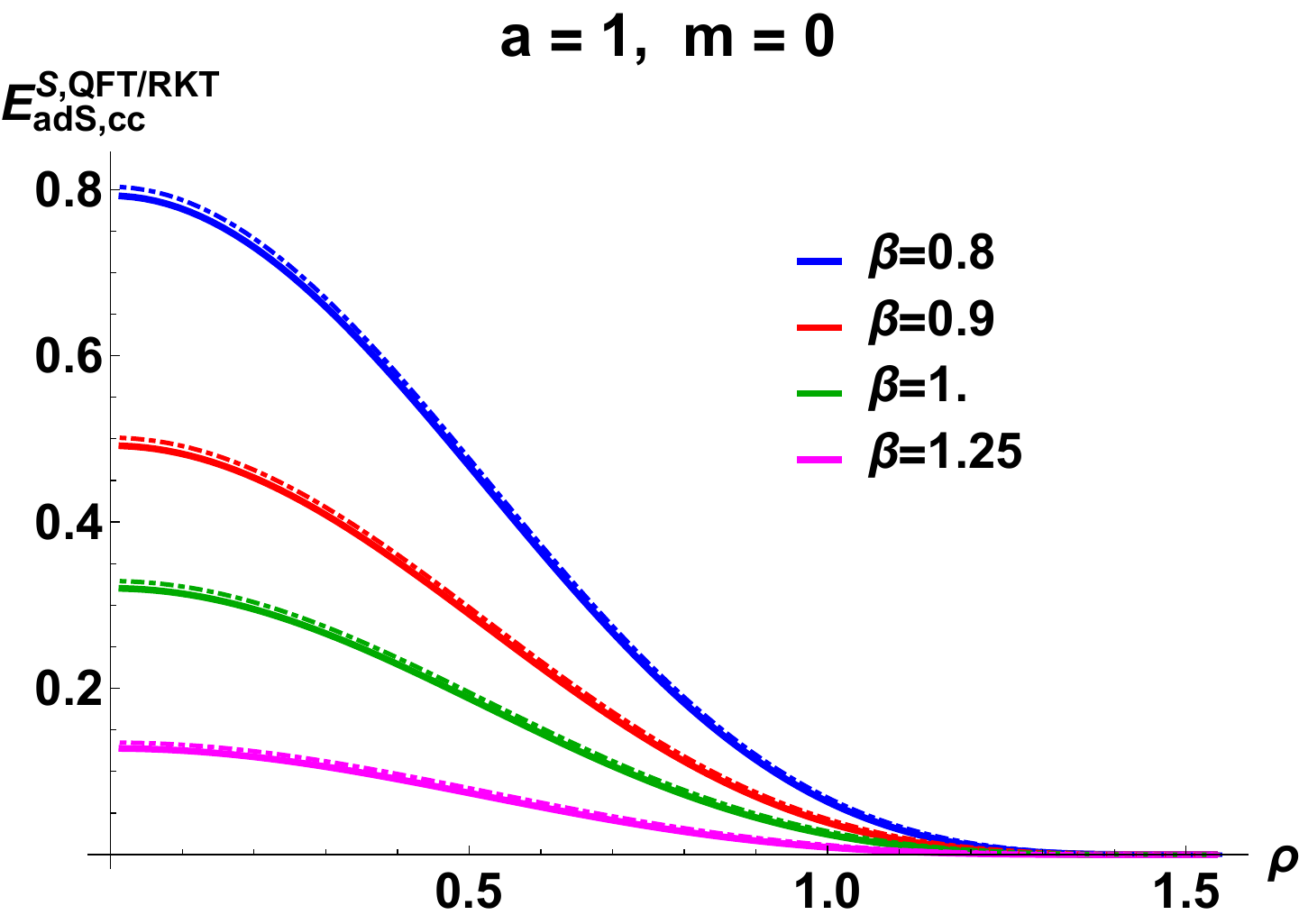} &
		\includegraphics[width=6.5cm]{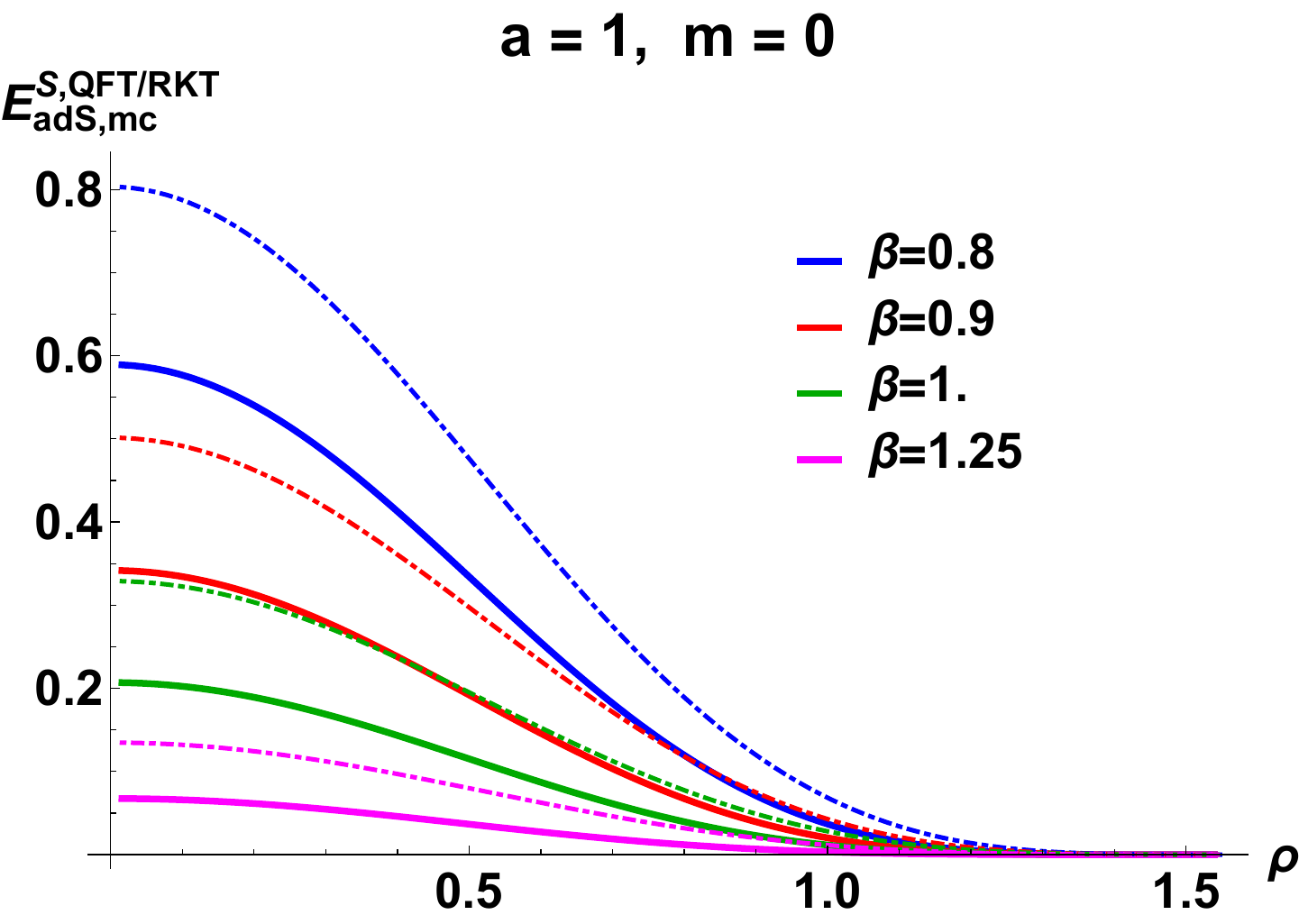} \\
		(a) & (b) \\
		\includegraphics[width=6.5cm]{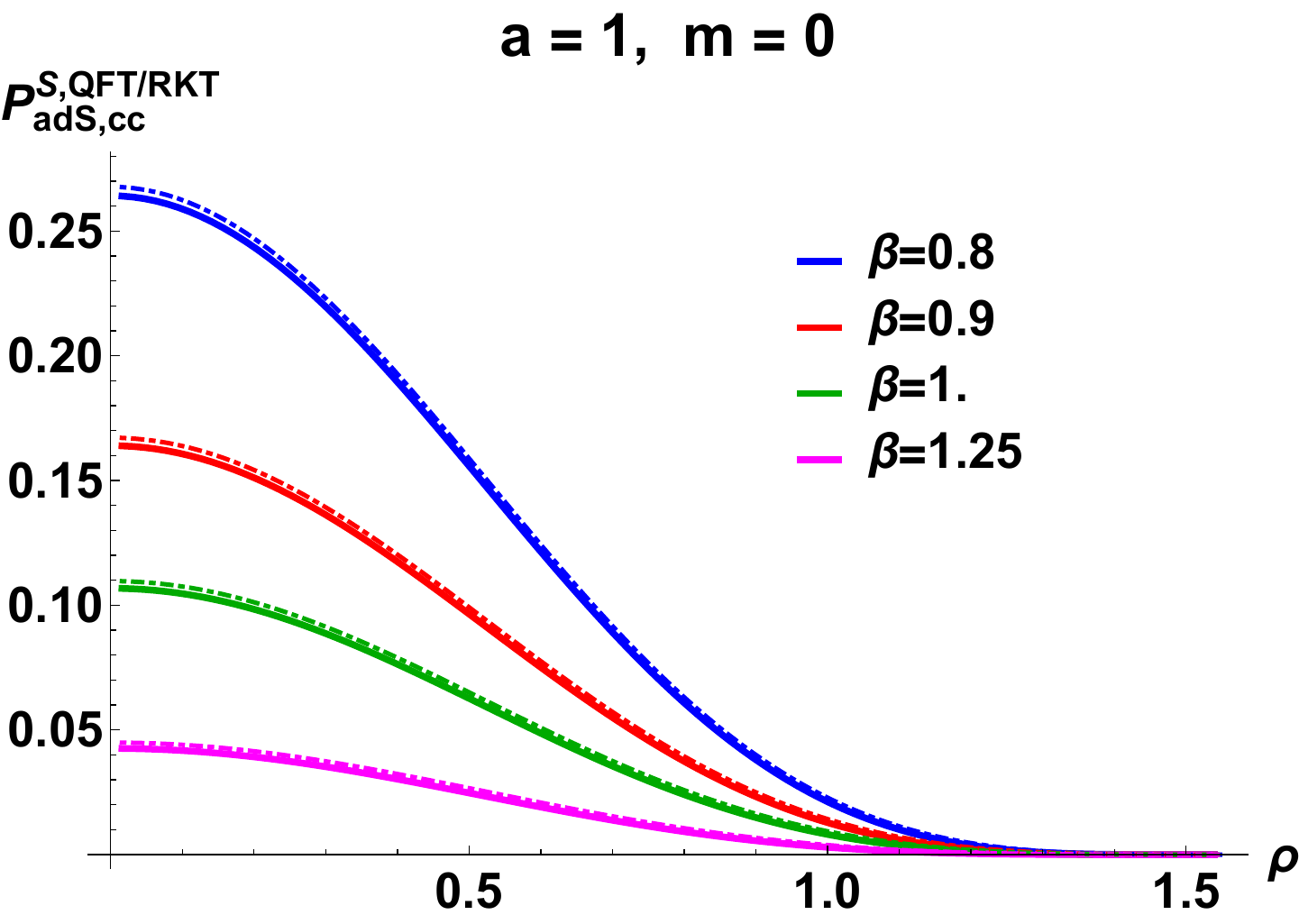} &
		\includegraphics[width=6.5cm]{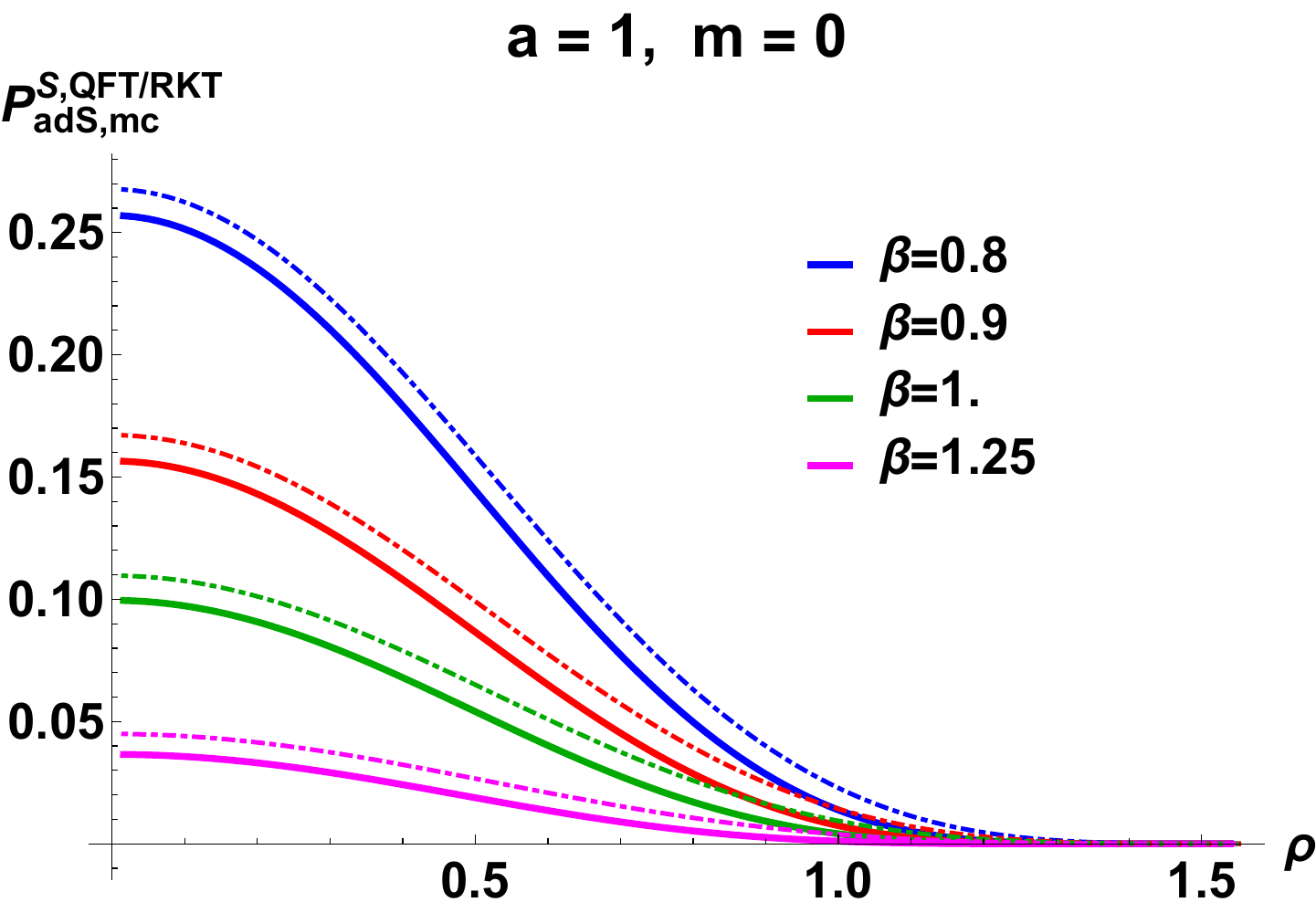} \\
		(c) & (d) \\
		\includegraphics[width=6.5cm]{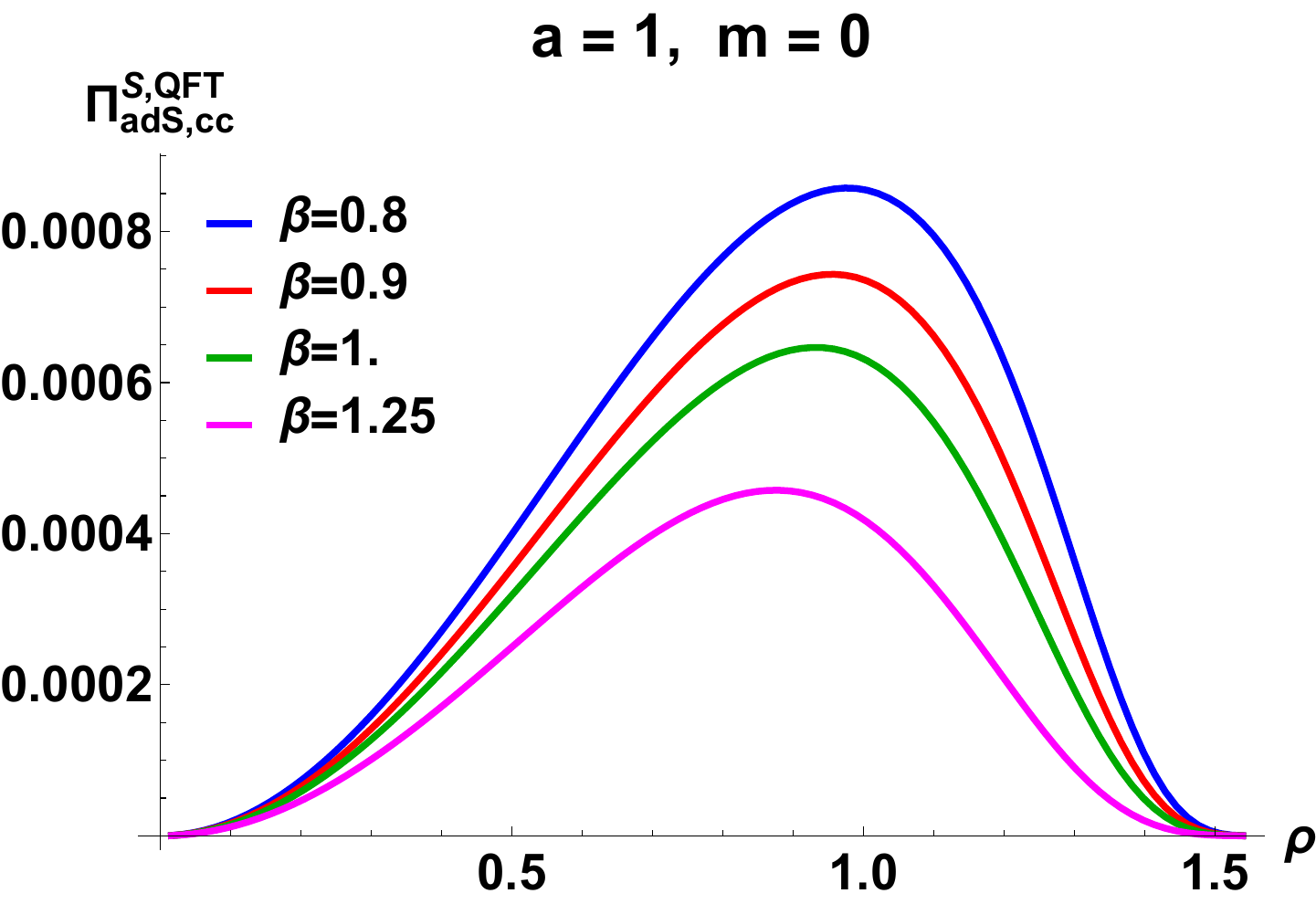} &
		\includegraphics[width=6.5cm]{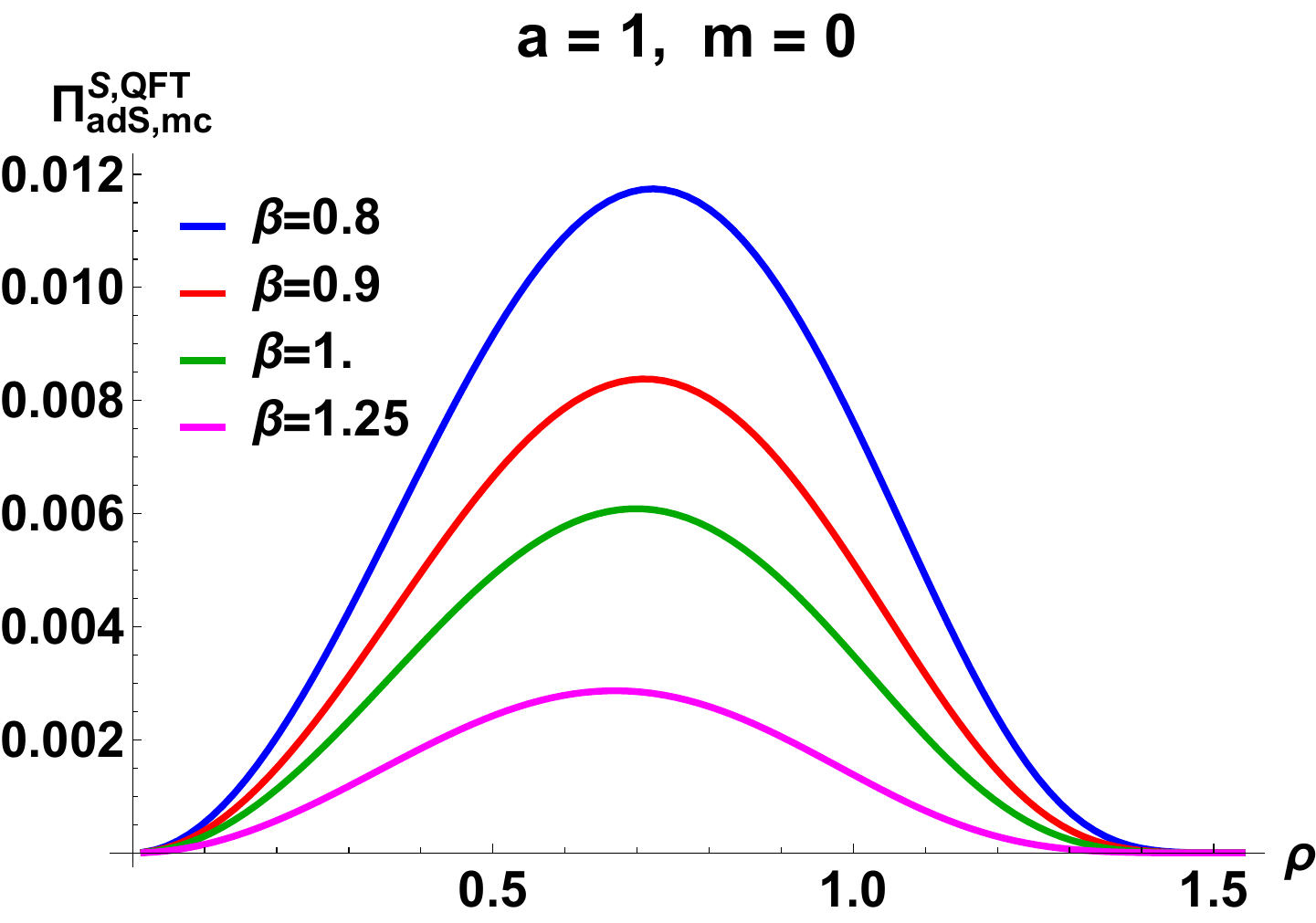} \\
		(e) & (f)
		\end{tabular}
	\end{center}
\caption{Energy density $E_{\mathrm {adS}}^{S,{\mathrm {QFT/RKT}}}$ (top), pressure $P_{\mathrm {adS}}^{S,{\mathrm {QFT/RKT}}}$ (middle) and pressure deviator $\Pi_{\mathrm {adS}}^{S,{\mathrm {QFT}}}$ (bottom) for massless scalars with conformal (left) and minimal (right) coupling to the adS scalar curvature. The adS radius of curvature is fixed to be $a=1$ and we consider a selection of values of the  inverse temperature $\beta $. All quantities are plotted as functions of the dimensionless radial coordinate $\rho$. Solid lines denote the QFT expressions (\ref{eq:E_thermal_cc}, \ref{eq:E_thermal_mc}, \ref{eq:QFTscalarP}), while dashed lines denote the corresponding RKT quantities (\ref{eq:masslessRKT}). In RKT, the pressure deviator is identically zero.  \label{fig:scalart1}}
\end{figure}

To explore this further, in Fig.~\ref{fig:scalart1} we plot the energy density $E_{\mathrm {adS}}^{S,{\mathrm {QFT}}}$, pressure $P_{\mathrm {adS}}^{S,{\mathrm {QFT}}}$ and pressure deviator $\Pi_{\mathrm {adS}}^{S,{\mathrm {QFT}}}$  for both conformally and minimally coupled massless scalar fields, and compare with the results (\ref{eq:masslessRKT}) from RKT. We fix the adS radius of curvature $a=1$ and consider a selection of values of the inverse temperature $\beta $.  All quantities are plotted as functions of the dimensionless radial coordinate $\rho $.

The overall shape of the profiles of the energy densities (\ref{eq:masslessRKT}, \ref{eq:E_thermal_cc}, \ref{eq:E_thermal_mc}) are similar, with the energy density tending to zero as $\rho \rightarrow \pi/2$ and the adS boundary is approached at similar rates in both QFT and RKT.  As expected, all quantities decrease as the inverse temperature $\beta $ increases.
For the conformally coupled scalar field (left-hand plots), the energy density and pressure in RKT and QFT are very similar, with the QFT quantities being slightly smaller than those in RKT. In contrast, for a minimally coupled scalar field we find that the energy density in QFT is significantly smaller than that in RKT.  The QFT pressure is also smaller than that in RKT, but the difference between the RKT and QFT pressures is smaller than the difference between the RKT and QFT energy densities.  

For both minimally and conformally coupled scalar fields, the pressure deviator is nonzero in QFT, unlike the situation in RKT where it vanishes identically.  The pressure deviator vanishes at the origin, has a maximum away from the origin, and is zero again on the adS boundary.  The pressure deviator is very small for a conformally coupled scalar field and an order of magnitude larger in the minimally coupled case (when it is roughly an order of magnitude smaller than the pressure).  Overall, quantum corrections are more significant for a minimally coupled massless scalar field
than for a conformally coupled massless scalar field at the same temperature.

\begin{figure}
	\begin{center}
		\begin{tabular}{cc}
			\includegraphics[width=6.5cm]{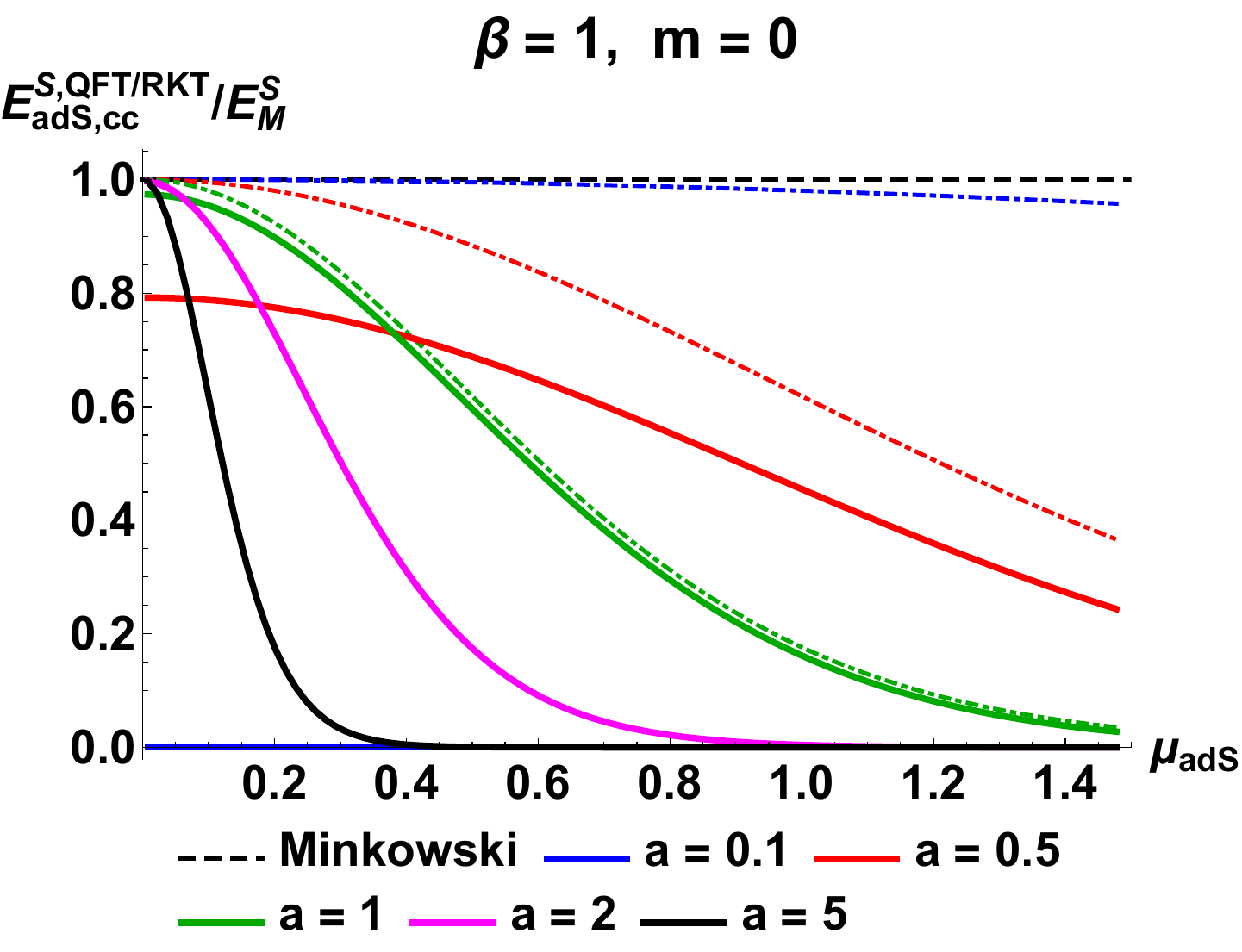} &
			\includegraphics[width=6.5cm]{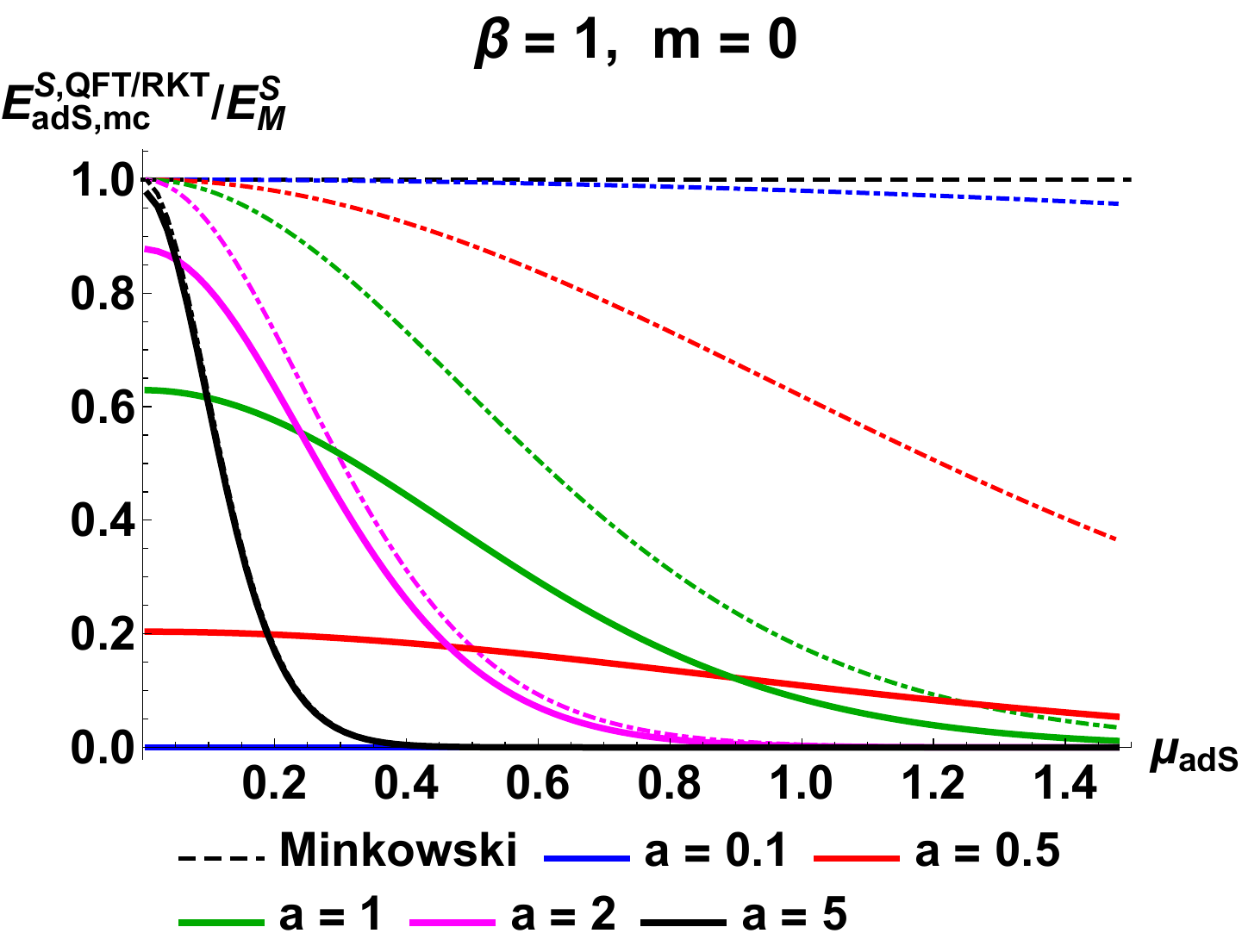} \\
			(a) & (b) \\ 
			\includegraphics[width=6.5cm]{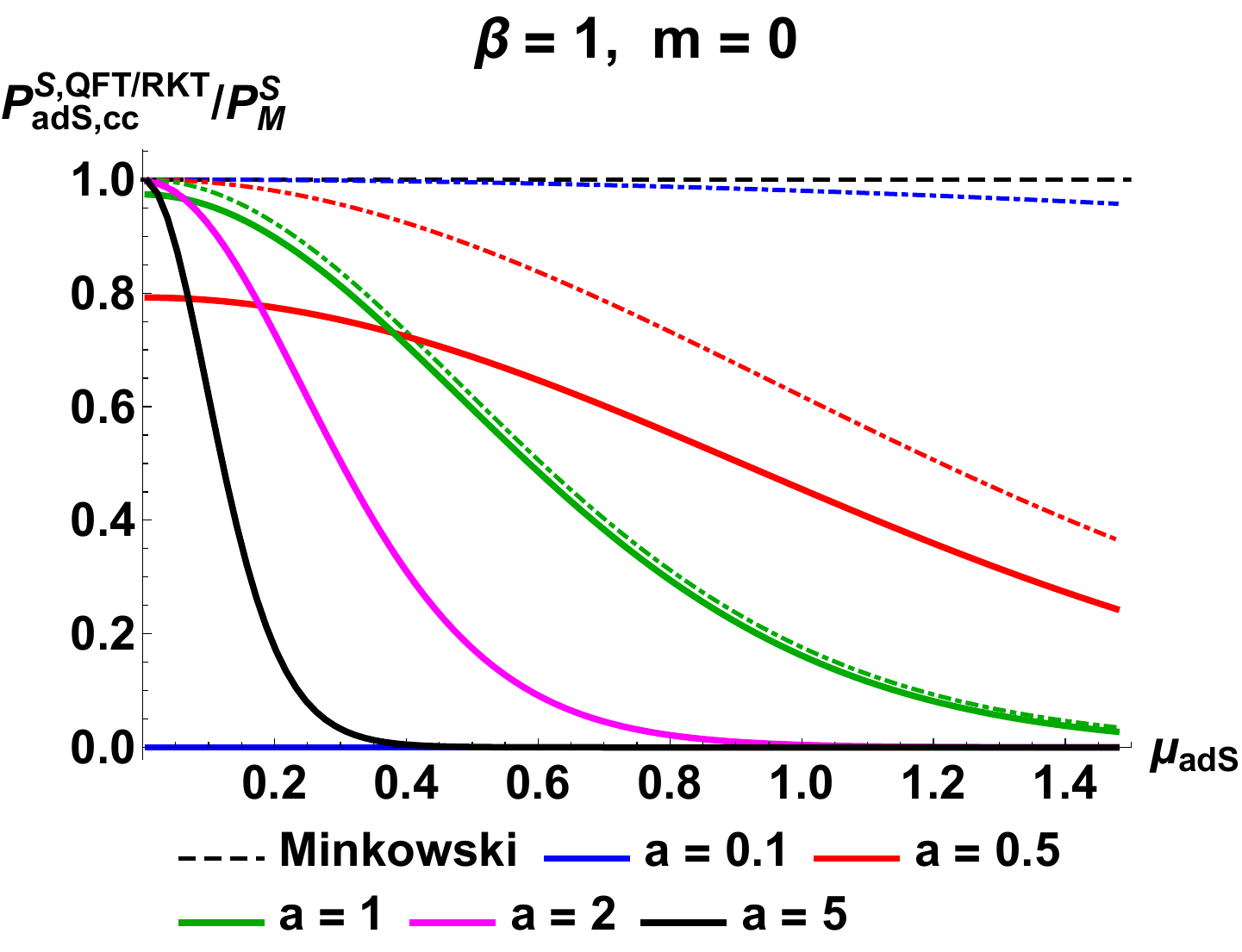} &
			\includegraphics[width=6.5cm]{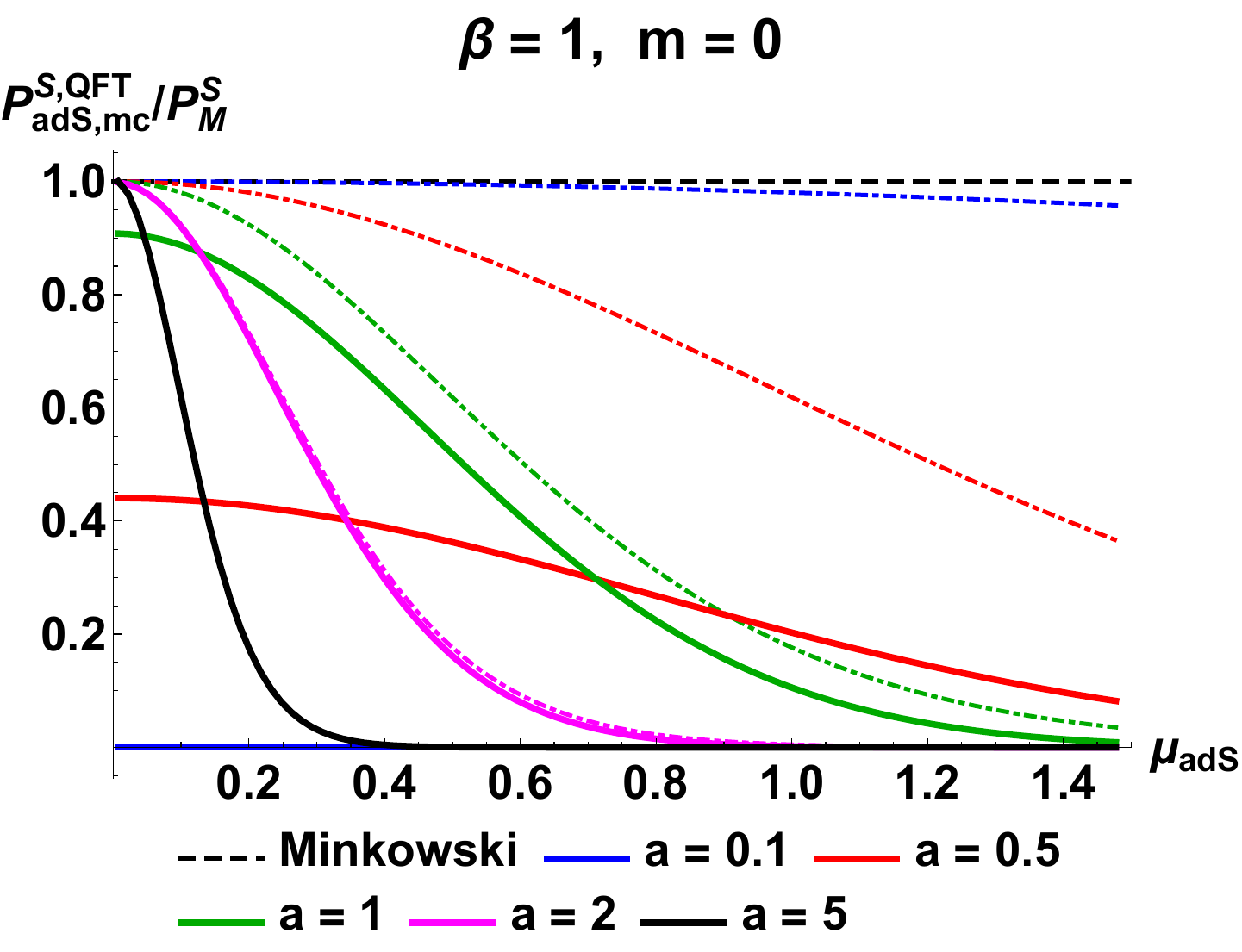} \\
			(c) & (d) \\ 
			\includegraphics[width=6.5cm]{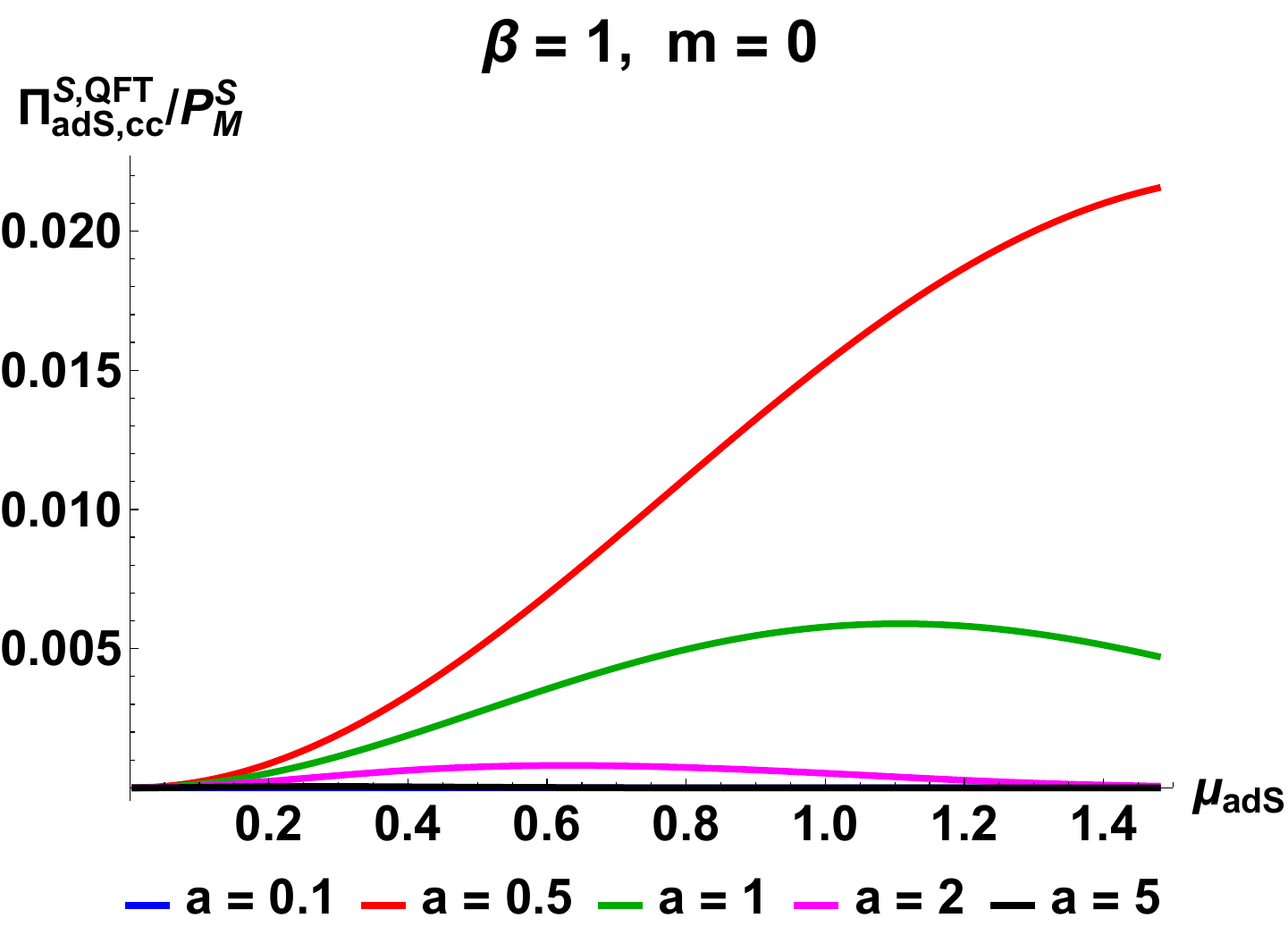} &
			\includegraphics[width=6.5cm]{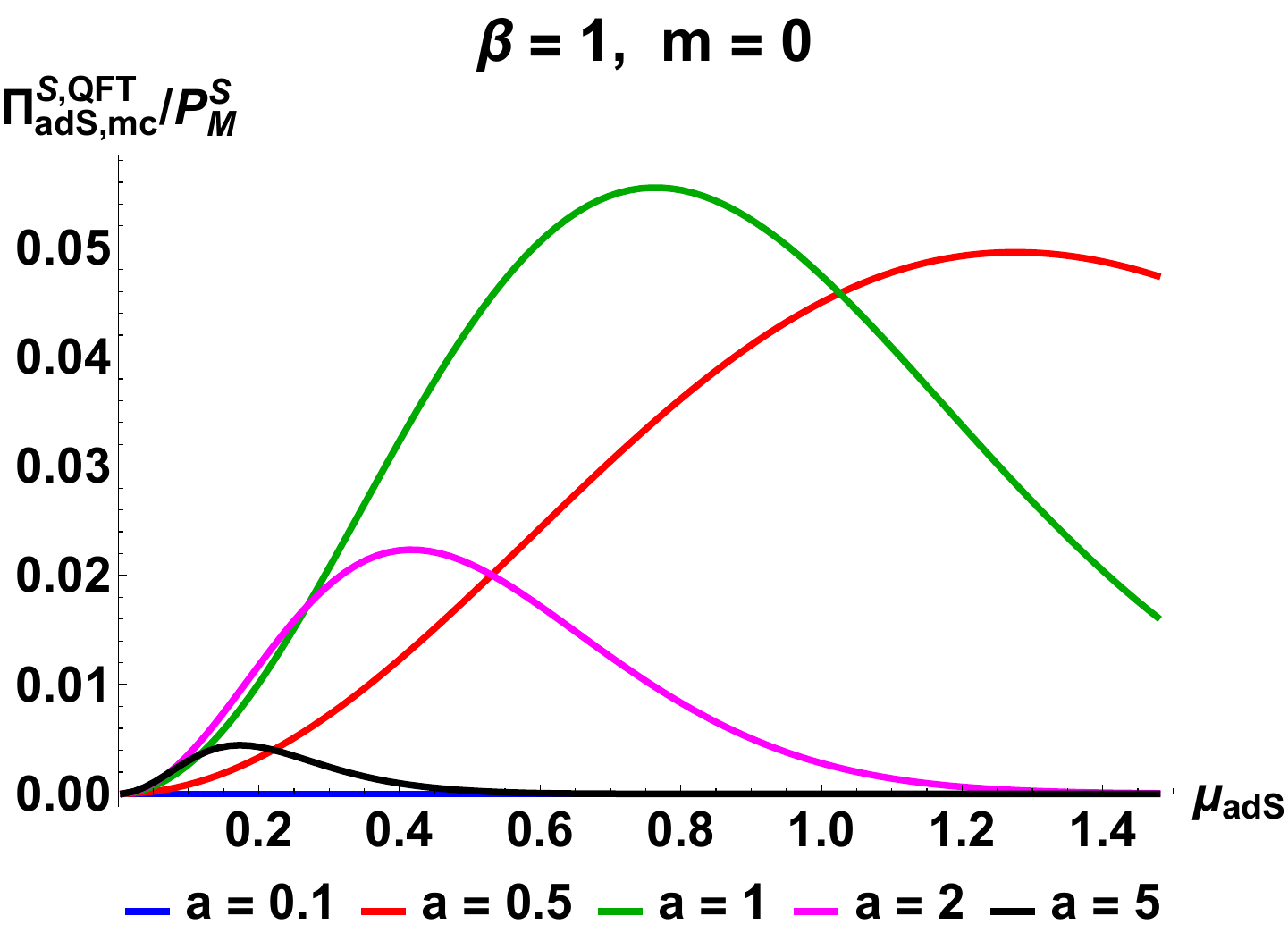} \\
			(e) & (f) 
			\end{tabular}
		\end{center}
		\caption{Energy density $E_{\mathrm {adS}}^{S,{\mathrm {QFT/RKT}}}$ (top), pressure $P_{\mathrm {adS}}^{S,{\mathrm {QFT/RKT}}}$ (middle) and pressure deviator $\Pi_{\mathrm {adS}}^{S,{\mathrm {QFT}}}$ (bottom) for massless scalars with conformal (left) and minimal (right) coupling to the adS scalar curvature. The inverse temperature is fixed to be  $\beta =1$, and all quantities are plotted as functions of the dimensionless geodetic distance $\mu _{\rm {adS}}$ (\ref{eq:muadS}). The energy density and pressure are divided by the corresponding Minkowski space-time values (\ref{eq:RKT_M}) and the pressure deviator is divided by the Minkowski space-time pressure. Solid lines denote QFT quantities, while dashed lines denote the corresponding RKT results.  In RKT, the pressure deviator is identically zero. \label{fig:scalart2}}
\end{figure}

In Fig.~\ref{fig:scalart1} we fixed the adS radius of curvature $a$ and varied the inverse temperature $\beta $. In Fig.~\ref{fig:scalart2}, we fix the inverse temperature $\beta =1$ and consider a selection of values of the adS inverse radius of curvature $a$.  We divide the energy density and pressure by the corresponding quantities in Minkowski space-time (\ref{eq:RKT_M}) and the pressure deviator by the Minkowski space-time pressure.  In Fig.~\ref{fig:scalart2}, we plot all quantities as functions of the dimensionless geodesic distance $\mu _{\rm {adS}}$ (\ref{eq:muadS}) defined in Sec.~\ref{sec:useful}.

In Fig.~\ref{fig:scalart2} we see that as the adS radius of curvature $a$ increases, and the Ricci scalar curvature decreases, the values of the energy density and pressure at the origin also increase.  For larger values of $a$, the profiles of the energy density and pressure rapidly tend to zero as $\mu _{\rm {adS}}$ increases. As $a$ decreases, the energy density and pressure decrease more slowly as $\mu _{\rm {adS}}$ increases. Quantum corrections become more important as $a$ decreases, and are larger for the minimally coupled field than for the conformally coupled field. 
For small values of $a$ ($a=0.1$ in the figure), both the energy density and pressure are negligible, and are strongly quenched by quantum corrections.

The pressure deviator (bottom plots in Fig.~\ref{fig:scalart2}) does not behave monotonically as $a$ varies. For both the minimally and conformally coupled cases, it is small when $a$ is large, then increases as $a$ decreases, reaches a maximum and then decreases again as $a$ decreases further.  As with the energy density and pressure, we see that the pressure deviator is negligible when $a$ is small. 

Comparing the left-hand and right-hand plots in Fig.~\ref{fig:scalart2}, we see marked differences in behaviour between the conformally and minimally coupled cases.  As in Fig.~\ref{fig:scalart1}, quantum corrections are larger in the minimally coupled case. The QFT energy density and pressure at the origin decrease more rapidly as $a$ decreases for minimal coupling compared to conformal coupling.  For conformal coupling, the pressure deviator reaches a maximum at larger values of $\mu _{\rm {adS}}$ than for minimal coupling.

\section{Quantum fermion field on adS space-time}
\label{sec:fermion}

We now consider a quantum fermion field $\Psi $ of mass $m$, which satisfies the Dirac equation
\begin{equation}
\left[ i\slashed{D}  -m \right] \Psi  =0,
\end{equation}
where $\slashed {D} =\gamma ^{\hat {\alpha }}D_{\hat {\alpha }}$ is the contraction between the spinor covariant derivative $D_{\hat {\alpha}} = \partial _{\hat {\alpha }} - \Gamma _{\hat {\alpha }}$ and the Minkowski Dirac matrices $\gamma ^{\hat {\alpha }}$.
The spin connection coefficients $\Gamma _{\hat {\alpha }}$ are given explicitly in \cite{Ambrus:2017cow}.
A fermion field minimally coupled to the space-time curvature is also conformally coupled; thus the two distinct cases considered for a quantum scalar field in the previous section reduce to one for the Dirac field studied in this section.

\subsection{Vacuum expectation values}
\label{sec:fermionvac}

As for the quantum scalar field, we first find the v.e.v.~of the SET for the quantum fermion field.
The vacuum Feynman Green's function satisfies the inhomogeneous Dirac equation
\begin{equation}
\left[ i\slashed{D}  -m \right]  S_{{\mathrm {vac}}}^{\mathrm {adS}}(x,x')= \left( -g \right)^{-\frac{1}{2}} \delta ^{4} (x-x') .
\label{eq:Diracinhomo}
\end{equation}
The vacuum Feynman Green's function takes a simple form due to the maximal symmetry of adS space-time \cite{Muck:1999mh,Camporesi:1992tm}:
\begin{equation}
	iS_{{\mathrm {vac}}}^{\mathrm {adS}}(x,x') = \left[{\mathcal {A}}_{F}(s_{\mathrm {adS}}) +{\mathcal {B}}_{F}(s_{\mathrm {adS}}) \slashed{n}\right]  \Lambda (x,x'). 
	\label{eq:SFadS}
\end{equation}
In the above expression, $\slashed{n}=\gamma ^{\hat{\alpha }}n_{\hat {\alpha }}$, where $n_{\hat {\alpha }}=\partial _{{\hat {\alpha }}} s_{\mathrm {adS}}$ is the tangent to the geodesic connecting the points $x$ and $x'$ at the point $x$, and $n_{\hat {\alpha }'}=\partial _{{\hat {\alpha }}'} s_{\mathrm {adS}}$ is the tangent at the point $x'$.
The bispinor $\Lambda (x,x')$ is the bispinor of parallel transport. This can be found in closed form on adS space-time. The expressions are lengthy so we do not reproduce them here -- the derivation of $\Lambda (x,x')$ and the final result are presented in the Appendix in \cite{Ambrus:2017cow}.
The remaining quantities in (\ref{eq:SFadS}) are the scalar functions ${\mathcal {A}}_{F}(s_{\mathrm {adS}})$ and ${\mathcal {B}}_{F}(s_{\mathrm {adS}})$, which depend only on the geodetic interval $s_{\mathrm {adS}}$. 
Their form is derived in a similar way to the scalar vacuum Feynman Green's function in Sec.~\ref{sec:scalarvac}, and the details can be found in \cite{Ambrus:2015mfa}.

Substituting the ansatz (\ref{eq:SFadS}) into the inhomogeneous Dirac equation (\ref{eq:Diracinhomo}) yields two coupled first order differential equations for ${\mathcal {A}}_{F}(s_{\mathrm {adS}})$ and ${\mathcal {B}}_{F}(s_{\mathrm {adS}})$, which can be combined into a single second order differential equation of hypergeometric form for ${\mathcal {A}}_{F}(s_{\mathrm {adS}})$.
The general solution for ${\mathcal {A}}_{F}(s_{\mathrm {adS}})$ then contains two arbitrary constants. The first is fixed by the requirement that the short-distance singularity structure of the vacuum Feynman Green's function must match that in Minkowski space-time; the second is fixed by requiring that the vacuum Feynman Green's function remains finite as $x'$ moves towards the adS boundary.
The final expression for ${\mathcal {A}}_{F}(s_{\mathrm {adS}})$ is then \cite{Ambrus:2015mfa}
\begin{subequations}
	\label{eq:AFBF}
\begin{eqnarray}
 {\mathcal {A}}_{F}(s_{\mathrm {adS}}) & = &  \frac {\Gamma \left( 2 + ma \right)}{16 \pi ^{\frac {3}{2}}a^{3} 4^{ma} \Gamma \left( \frac {1}{2} + ma \right) }
\cos \left( \frac {s_{\mathrm {adS}}}{2a} \right) \left[
-\sin^{2} \left( \frac {s_{\mathrm {adS}}}{2a} \right) \right] ^{-2-ma}
\nonumber \\ & & \times
{}_{2}F_{1}\left[  1+ma, 2+ma; 1+2ma ; {\rm cosec} ^{2} \left( \frac {s_{\mathrm {adS}}}{2a} \right) \right],
\label{eq:calA}
\end{eqnarray}
from which ${\mathcal {B}}_{F}(s_{\mathrm {adS}})$ is determined to be \cite{Ambrus:2015mfa}
\begin{eqnarray}
{\mathcal {B}}_F(s_{\mathrm {adS}})  & = & \frac {i\Gamma \left( 2 + ma \right)}{16 \pi ^{\frac {3}{2}}a^{3} 4^{ma} \Gamma \left( \frac {1}{2} + ma \right) }
\sin \left( \frac {s_{\mathrm {adS}}}{2a} \right) \left[
-\sin^{2} \left( \frac {s_{\mathrm {adS}}}{2a} \right) \right] ^{-2-ma}
\nonumber \\
 & & \times 
{}_{2}F_{1}\left[ ma, 2+ma; 1+2ma ; {\rm cosec} ^{2} \left( \frac {s_{\mathrm {adS}}}{2a} \right) \right] .
\end{eqnarray}
\end{subequations}
When the fermion field is massless, $m=0$, the expressions (\ref{eq:AFBF}) simplify considerably:
\begin{eqnarray}
 \left. {\mathcal {A}}_F(s_{\mathrm {adS}}) \right\rfloor _{m=0} & = & \frac{1}{16\pi^2a^{3}} \left[\cos\left( \frac{s_{\mathrm {adS}}}{2a} \right) \right] ^{-3}, 
 \nonumber \\
\left. {\mathcal {B}}_F(s_{\mathrm {adS}}) \right\rfloor_{m=0} & = & \frac{i}{16\pi^2a^{3}} \left[\sin \left( \frac{s_{\mathrm {adS}}}{2a}\right) \right] ^{-3} .
\end{eqnarray}

The v.e.v.~of the SET is computed from the vacuum Feynman Green's function (\ref{eq:SFadS}) by applying the curved space-time stress energy tensor operator:
\begin{eqnarray}
\braket{T_{{\hat {\alpha }}{\hat {\nu}}}}^{{\mathrm {F}},\mathrm {adS}}_{\mathrm {vac}}
& = & \frac{i}{2}\lim_{x'\rightarrow x}  
{\mathrm {Tr}} \ \left\{
\left[ \gamma_{({\hat {\alpha }}} D_{{\hat {\nu}} )} iS_{{\mathrm {vac}}}^{\mathrm {adS}}(x, x')
\right.\right. \nonumber \\ & & \left.\left. -
D_{{\hat {\alpha }}'} [iS_{{\mathrm {vac}}}^{\mathrm {adS}}(x,x')]
\gamma_{{\hat {\nu }}'} g^{{\hat {\alpha }}'}{}_{({\hat {\alpha }}}
g^{{\hat {\nu }}'}{}_{{\hat {\nu }})}\right] \Lambda(x',x) 
 \right\} .
\end{eqnarray}
As in the scalar case, this procedure yields an infinite quantity and we apply Hadamard renormalization to give a finite v.e.v.~for the SET.

Since, as seen in Sec.~\ref{sec:scalarvac}, we have a well-defined prescription for Hadamard renormalization for a quantum scalar field, in the fermion case we begin by defining an auxiliary propagator ${\mathcal {G}}^{\mathrm {adS}}_{\mathrm {vac}}(x,x')$, which is related to the fermion vacuum Feynman Green's function $S_{{\mathrm {vac}}}^{\mathrm {adS}}(x, x')$ (\ref{eq:SFadS}) by \cite{Najmi:1985yi}
\begin{equation}
iS_{{\mathrm {vac}}}^{\mathrm {adS}}(x, x') = \left( i \slashed{D} + m \right) i{\mathcal{G}}^{\mathrm {adS}}_{\mathrm {vac}}(x,x').
\label{eq:calg_def}
\end{equation}
This is reminiscent of the relationship (\ref{eq:fermionvacM}) between the scalar and fermion vacuum Feynman Green's functions on Minkowski space-time. However, the auxiliary propagator  ${\mathcal{G}}^{\mathrm {adS}}_{\mathrm {vac}}(x,x')$,  like the fermion vacuum  Feynman Green's function $S_{{\mathrm {vac}}}^{\mathrm {adS}}(x, x')$, is a bispinor. This is not immediately apparent in the Minkowski case (\ref{eq:fermionvacM}) when the bispinor of parallel transport $\Lambda (x,x')$ is the identity matrix.

The inhomogeneous Dirac equation (\ref{eq:Diracinhomo}) implies that the auxiliary propagator ${\mathcal{G}}^{\mathrm {adS}}_{\mathrm {vac}}(x,x')$ satisfies a Klein-Gordon-like equation \cite{Najmi:1985yi}
\begin{equation}
\left[\Box _{x} - \tfrac{1}{4} R - m^2\right] {\mathcal{G}}^{\mathrm {adS}}_{\mathrm {vac}}(x,x') = \left(-g\right) ^{-\frac{1}{2}} \delta^4(x-x'),
\label{eq:calG_eq}
\end{equation}
where, because ${\mathcal{G}}^{\mathrm {adS}}_{\mathrm {vac}}(x,x')$ is a bispinor, the $\Box _{x}$ operator is now a combination of spinor covariant derivatives $\Box _{x} = g^{{\hat {\alpha }}{\hat {\rho}}} D_{\hat {\alpha }}D_{\hat {\rho}}$.
From the form (\ref{eq:SFadS}) of the fermion vacuum Feynman Green's function, the auxiliary propagator can be shown to be proportional to the bispinor of parallel transport \cite{Ambrus:2015mfa}
\begin{equation}
i{\mathcal{G}}^{\mathrm {adS}}_{\mathrm {vac}}(x,x') = \frac{{\mathcal {A}}_{F}(s_{\mathrm {adS}})}{m} \Lambda(x,x'),
\label{eq:calG_form}
\end{equation}
where ${\mathcal {A}}_{F}(s_{\mathrm {adS}})$ is the scalar function (\ref{eq:calA}).  In analogy to the scalar Hadamard parametrix (\ref{eq:scalarH}), the Hadamard parametrix for the auxiliary propagator ${\mathcal{G}}^{\mathrm {adS}}_{\mathrm {vac}}(x,x')$ is \cite{Najmi:1985yi,Dappiaggi:2009xj}
\begin{equation}
i{\mathcal{G}}_{\rm {Had}}(x,x') = \frac{1}{8\pi^2} \left[\frac{{\mathcal {U}}(x,x')}{\sigma} +
{\mathcal {V}}(x,x')\,\ln(M^2\sigma)\right] ,
\label{eq:calG_Hadamard}
\end{equation}
where $M$ is an arbitrary mass renormalization scale.
The quantities ${\mathcal {U}}(x,x')$ and ${\mathcal {V}}(x,x')$ are bispinors which are regular in the limit $x'\rightarrow x$.
Since we are working in four space-time dimensions, the bispinor ${\mathcal {U}}(x,x')$ takes the simple form \cite{Najmi:1985yi} (cf.~(\ref{eq:scalarH}))
\begin{equation}
{\mathcal {U}}(x,x') = \Delta (x,x')^{\frac{1}{2}}\Lambda (x,x'),
\end{equation}
where $\Delta (x,x')$ is the van-Vleck-Morette determinant (\ref{eq:VVMD}).

The bispinor ${\mathcal {V}}(x,x')$ satisfies the homogeneous version of (\ref{eq:calG_eq}). 
Writing ${\mathcal {V}}(x,x')$ in a form similar to (\ref{eq:calG_form})
\begin{equation}
{\mathcal {V}}(x,x') = \frac{{\mathcal {A}}_{V}(s_{\mathrm{adS}})}{m}\Lambda (x,x'),
\end{equation}
we find that ${\mathcal {A}}_{V}(s_{\mathrm {adS}})$ satisfies the same hypergeometric differential equation as ${\mathcal {A}}_{F}(s_{\mathrm {adS}})$, but with different boundary conditions (${\mathcal {A}}_{V}(s_{\mathrm {adS}})$ is regular as $s_{\mathrm {adS}}\rightarrow 0$, but ${\mathcal {A}}_{F}(s_{\mathrm {adS}})$ is singular in this limit).
The appropriate solution of the differential equation is
\begin{equation}
{\mathcal {A}}_V(s_{\mathrm {adS}}) = m{\tilde {\mathcal{C}}}_{V}\cos\left( \frac{s_{\mathrm{adS}}}{2a} \right)  {}_2F_1\left[  2-ma,2+ma;2; \sin ^{2}\left(\frac{s_{\mathrm {adS}}}{2a}\right)\right] ,
\label{eq:alpha_V_sol}
\end{equation}
where ${\tilde {\mathcal {C}}}_{V}$ is an arbitrary constant.  This is fixed using the boundary condition on ${\mathcal {V}}(x,x')$ analogous to (\ref{eq:CVBC}), which is
\begin{equation}
{\tilde {{\mathcal {C}}}}_{V} = \lim_{s_{\mathrm {adS}}\rightarrow 0} \left[
\frac{1}{2}\left( m^{2}+\frac{1}{4} R - \Box _{x}\right) \Delta (x,x')^{\frac{1}{2}}\Lambda (x,x') 
\right] = \frac{1}{2a^{2}}\left( m^{2}a^{2}-1\right) .
\end{equation}
Having determined the Hadamard parametrix (\ref{eq:calG_Hadamard}) for the auxiliary propagator, the corresponding Hadamard parametrix for the fermion Feynman Green's function is then
\begin{equation}
iS_{\rm {Had}}(x,x') = \left( i\slashed {D} + m  \right) i {\mathcal {G}}_{\rm {Had}} (x,x').
\label{eq:fermion_Hadamard}
\end{equation}
As in the scalar field case, the Hadamard parametrix (\ref{eq:fermion_Hadamard}) for the spinor Feynman Green's function is purely geometric and independent of the quantum state under consideration.
To compute the v.e.v.~of the SET, one would therefore expect to subtract the Hadamard parametrix (\ref{eq:fermion_Hadamard}) from the vacuum fermion Feynman Green's function (\ref{eq:SFadS}) and then apply the SET operator, before bringing the points together: 
\begin{eqnarray}
\braket{T_{{\hat {\alpha }}{\hat {\nu}}}}^{{\mathrm {F}},\mathrm {adS}}_{\mathrm {vac}}
& = & \frac{i}{2}\lim_{x'\rightarrow x}  
{\mathrm {Tr}} \ \left\{
\left[ \gamma_{({\hat {\alpha }}} D_{{\hat {\nu}} )} i\left( S_{{\mathrm {vac}}}^{\mathrm {adS}}(x, x') - S_{\mathrm {Had}}(x,x')\right) 
\right.\right. \nonumber \\ & & \left.\left. \hspace{-0.5cm}-
D_{{\hat {\alpha }}'} [i\left( S_{{\mathrm {vac}}}^{\mathrm {adS}}(x,x')- S_{\mathrm {Had}}(x,x') \right) ]
\gamma_{{\hat {\nu }}'} g^{{\hat {\alpha }}'}{}_{({\hat {\alpha }}}
g^{{\hat {\nu }}'}{}_{{\hat {\nu }})}\right] \Lambda(x',x) 
\right\} .
\label{eq:fermion_SET_uncons}
\end{eqnarray}
There is however an additional complication \cite{Dappiaggi:2009xj}.  In the Hadamard renormalization of the scalar Feynman Green's function, the Hadamard parametrix $G_{\rm {Had}}(x,x')$ (\ref{eq:scalarH}) is a solution of the inhomogeneous Klein-Gordon equation (\ref{eq:KGinhomo}). Since the vacuum Feynman Green's function $G_{\rm {vac}}^{\rm {adS}}(x,x')$ also satisfies the inhomogeneous Klein-Gordon equation, it follows that the regularized propagator $G_{\rm {vac}}^{\rm {adS}}(x,x')-G_{\rm {Had}}(x,x')$ is a solution of the homogeneous Klein-Gordon equation, and applying the SET operator (\ref{eq:scalar_SET}) yields a conserved SET, as required.
However the Hadamard parametrix (\ref{eq:fermion_Hadamard}) is not a solution of the inhomogeneous Dirac equation \cite{Dappiaggi:2009xj}. This means that the SET computed using (\ref{eq:fermion_SET_uncons}) will not be conserved. This is resolved \cite{Dappiaggi:2009xj} by adding to the classical fermion SET a term proportional to the Dirac Lagrangian. This will vanish for classical solutions of the Dirac equation. For the quantum fermion field, the corresponding modified SET operator is \cite{Dappiaggi:2009xj}
\begin{eqnarray}
\braket{T_{{\hat {\alpha }}{\hat {\nu}}}}^{{\mathrm {F}},\mathrm {adS}}_{\mathrm {vac}}
& = & \frac{i}{2}\lim_{x'\rightarrow x}  
{\mathrm {Tr}} \ \left\{
\left[ \gamma_{({\hat {\alpha }}} D_{{\hat {\nu }} )} i\left( S_{{\mathrm {vac}}}^{\mathrm {adS}}(x, x') - S_{\mathrm {Had}}(x,x')\right) 
\right.\right. \nonumber \\ & & \left.\left. -
D_{{\hat {\alpha }}'} [i\left( S_{{\mathrm {vac}}}^{\mathrm {adS}}(x,x')- S_{\mathrm {Had}}(x,x') \right) ]
\gamma_{{\hat {\nu }}'} g^{{\hat {\alpha }}'}{}_{({\hat {\alpha }}}
g^{{\hat {\nu }}'}{}_{{\hat {\nu }})}
\right.\right. \nonumber \\ & & 
\left.\left.
- \frac {1}{12} g_{\hat {\alpha }{\hat {\nu }}} 
\left\{
 \gamma^{{\hat {\sigma }}} D_{{\hat {\sigma }} } i\left( S_{{\mathrm {vac}}}^{\mathrm {adS}}(x, x') - S_{\mathrm {Had}}(x,x')\right) 
\right. \right.\right. \nonumber \\ & & \left.\left. \left.
-D_{{\hat {\lambda }}'} [i\left( S_{{\mathrm {vac}}}^{\mathrm {adS}}(x,x')- S_{\mathrm {Had}}(x,x') \right) ]
\gamma_{{\hat {\sigma }}'} g^{{\hat {\lambda  '}}}{}_{({\hat {\lambda  }}}
g^{{\hat {\sigma}}'}{}_{{\hat {\sigma }})} g^{{\hat {\lambda }}{\hat \sigma }}
\right. \right. \right. \nonumber \\ & & \left. \left. \left.
-2mi\left( S_{\mathrm {vac}}^{\mathrm {adS}} (x,x')  - S_{\mathrm {Had}}(x,x') \right) 
\right\}
\right] \Lambda(x',x) 
\right\} ,
\label{eq:fermion_SET_cons}
\end{eqnarray}
which now yields a conserved SET, as required.

The v.e.v.~of the SET can then be explicitly computed, giving a compact expression \cite{Ambrus:2015mfa}:
\begin{eqnarray}
\braket{T_{\hat{\alpha}}^{\hat{\nu}}}^{{\mathrm {F}},\mathrm {adS}}_{\mathrm {vac}}
& = & 
\left\{ -\frac{1}{16\pi ^{2}a^{4}}\left(
\frac{11}{60} + ma - \frac{7m^2a^{2}}{6} - m^3a^{3} + \frac{3m^4a^{4}}{2}\right)
\right. \nonumber \\ & & \left.
+ \frac{m^2(m^2a^{2} - 1)}{2\pi^2 a^{2}} \Upsilon _{F}  \right\} \delta _{\hat {\alpha }}^{\hat {\nu }},
\label{eq:vevFadS}
\end{eqnarray}
where
\begin{equation}
\Upsilon _{F} = \psi (ma) + C - \ln \left( {\sqrt {2}}Ma \right) . 
\end{equation}
As in the scalar case, the expression (\ref{eq:vevFadS}) agrees with that obtained by $\zeta$-function regularization \cite{Camporesi:1992wn}.

The tetrad components of the v.e.v.~of the SET (\ref{eq:vevFadS}) are constants for all values of the fermion mass $m$, as was also found in Sec.~\ref{sec:scalarvac} for scalar fields. In general, the v.e.v.~depends on the arbitrary mass renormalization scale $M$, unless either $ma=0$ or $ma=1$.
In the massless case, the v.e.v.~of the SET simplifies to 
\begin{equation}
\left. \braket{T_{\hat{\alpha}}^{\hat{\nu}}}^{{\mathrm {F}},\mathrm {adS}}_{\mathrm {vac}} \right\rfloor _{m=0} =-\frac{11}{960\pi ^{2}a^{4}}\delta _{\hat {\alpha }}^{\hat {\nu }}.
\label{eq:fermion_vev}
\end{equation} 

\subsection{Thermal expectation values}
\label{sec:fermionT}

The computation of the t.e.v.~of the SET for fermions on adS follows the same steps as those for the scalar case in Sec.~\ref{sec:scalarT}.
We start with the thermal fermion Feynman Green's function, defined in analogy to (\ref{eq:fermion_thermal_M}) on Minkowski space-time:
\begin{equation}
S_{\beta }^{\rm {adS}}(x,x') =  
\sum _{j=-\infty }^{\infty } (-1)^{j}S_{\rm {vac}}^{\rm {adS}} (\tau +ij{\bar {\beta}},{\bm {x}}; \tau ',\bm {x'}) ,
\label{eq:fermion_thermal_ads_S}
\end{equation}
where ${\bar {\beta}}$ is the dimensionless inverse temperature (\ref{eq:betabar}).
Following the same approach as that for a quantum scalar field in Sec.~\ref{sec:scalarT}, we consider the difference between the t.e.v.~and the v.e.v.~of the SET, which does not require renormalization, and is given by
\begin{eqnarray}
\braket{: T_{{\hat {\alpha }}{\hat {\nu }}} :}^{{\mathrm {F}},\mathrm {adS}}_{\beta }
& = & \frac{i}{2}\lim_{x'\rightarrow x}  
{\mathrm {Tr}} \ \left\{
\left[ \gamma_{({\hat {\alpha }}} D_{{\hat {\nu }} )} i\left( S_{{\mathrm {vac}}}^{\beta }(x, x') - S^{\mathrm {adS}}_{\mathrm {vac}}(x,x')\right) 
\right.\right. \nonumber \\ & & \left.\left. 
\hspace{-0.9cm}-
D_{{\hat {\alpha }}'} [i\left( S_{\beta }^{\mathrm {adS}}(x,x')- S^{\mathrm {adS}}_{\mathrm {vac}}(x,x') \right) ]
\gamma_{{\hat {\nu }}'} g^{{\hat {\alpha }}'}{}_{({\hat {\alpha }}}
g^{{\hat {\nu }}'}{}_{{\hat {\nu }})}\right] \Lambda(x',x) 
\right\} .
\label{eq:fermion_thermal_adS}
\end{eqnarray}
Note that both $S_{\beta}^{\mathrm{adS}}(x,x')$ and $S_{\mathrm{vac}}^{\mathrm{adS}}(x,x')$ are solutions of the inhomogeneous Dirac equation (\ref{eq:Diracinhomo}), and therefore we do not need to add the additional terms to the stress-energy tensor operator which are required for the computation of the v.e.v.~(these additional terms would vanish anyway in this case).

The details of the derivation of the components of (\ref{eq:fermion_thermal_adS}) can be found in \cite{Ambrus:2017cow} for a fermion field of mass $m$. 
Here, for simplicity, we restrict our attention to the massless case, when the components of   (\ref{eq:fermion_thermal_adS}) take 
the perfect fluid form (\ref{eq:Landau}) with 
\begin{eqnarray}
\frac{1}{3} E_{\mathrm{adS}}^{F,{\mathrm {QFT}}}(\beta ) = P_{\mathrm{adS}}^{F,{\mathrm {QFT}}}(\beta ) 
 & = & \frac{\cos ^{4}\rho} {4\pi ^{2}a^{4}} \sum _{j=1}^{\infty } (-1)^{j-1} \frac{\cosh \left( \frac{j{\bar {\beta}}}{2} \right)}{\sinh ^{4}\left( \frac{j{\bar {\beta}}}{2}\right)} ,
\nonumber \\ 
\Pi _{\mathrm{adS}}^{F,{\mathrm {QFT}}}(\beta ) & = & 0. 
\label{eq:QFTFE}
\end{eqnarray}
We immediately note a difference compared to the scalar field case, namely that the pressure deviator $\Pi _{\mathrm{adS}}^{F,{\mathrm {QFT}}}(\beta )$ vanishes identically.  Interestingly, this turns out to be the case, not only for massless fermions, but also for fermions with nonzero mass \cite{Ambrus:2017cow}.

\begin{figure}
	\begin{center}
		\includegraphics[width=10cm]{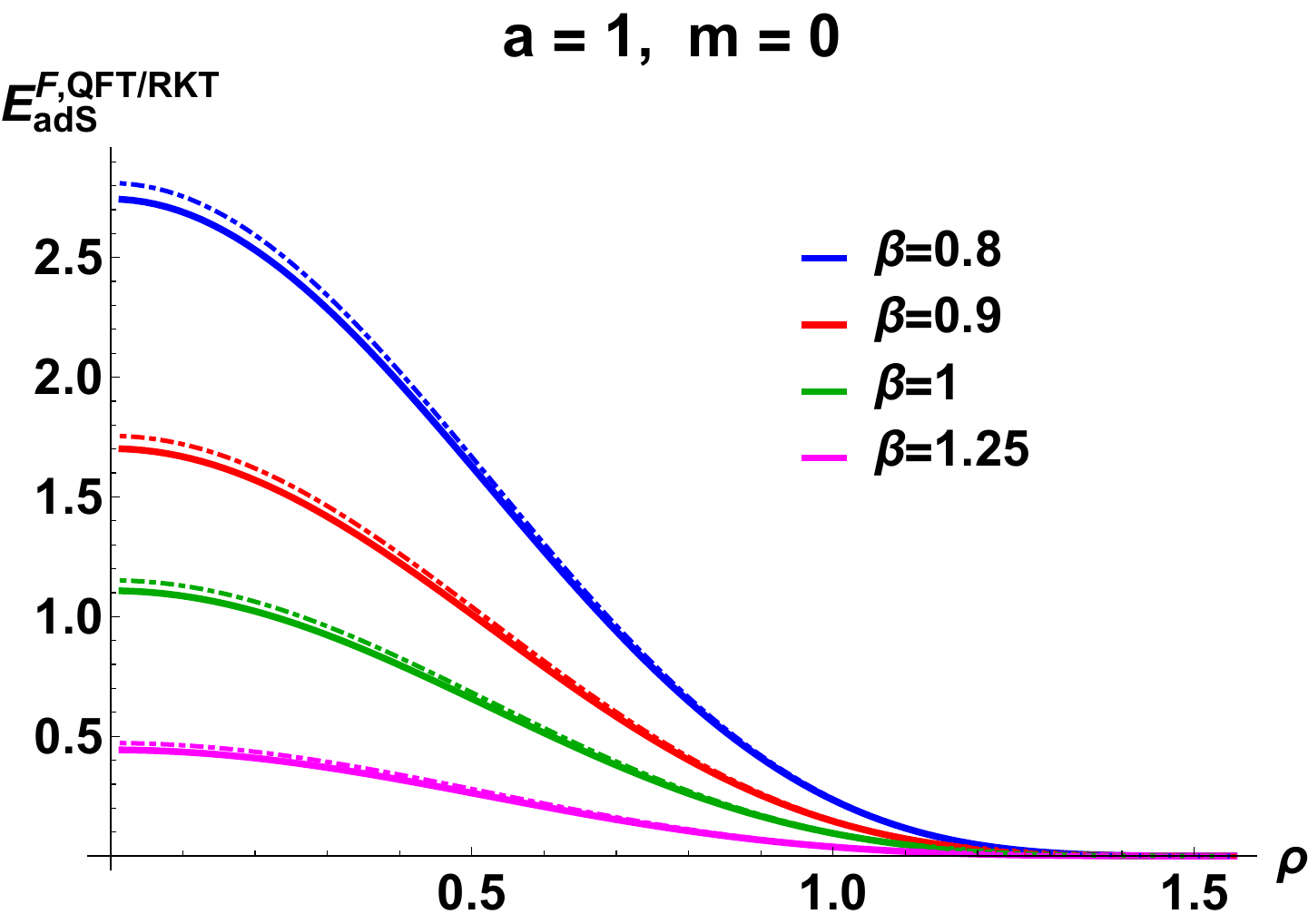}
		\caption{Energy density $E_{\mathrm {adS}}^{F,{\mathrm {QFT/RKT}}}$ for massless fermions with adS radius of curvature $a=1$ and a selection of values of the inverse temperature $\beta $, plotted as functions of the dimensionless radial coordinate $\rho$. Solid lines denotes the QFT expressions (\ref{eq:QFTFE}), while dashed lines denote the corresponding RKT quantities (\ref{eq:masslessRKT}). \label{fig:fermiont1}}
	\end{center}
\end{figure}

In Fig.~\ref{fig:fermiont1} we plot the profiles of the energy density for massless fermions with the adS radius of curvature fixed to be $a=1$ and a selection of values of the inverse temperature $\beta $. Since the pressure is one third the energy density, we have not presented a plot of the pressure.  We see that the RKT energy density (\ref{eq:masslessRKT}) is a very good approximation to the QFT energy density (\ref{eq:QFTFE}) for these values of $\beta $ and $a$.  In \cite{Ambrus:2017cow} it was found that the RKT quantities very closely approach those derived in QFT when the temperature is large, in which case the fermions behave essentially classically.  The energy profiles are very similar in overall shape to those seen in Fig.~\ref{fig:scalart1} for scalar fields:~there is a maximum at the origin, and the energy density decreases to zero as $\rho \rightarrow \pi/2$ and the adS boundary is approached.

\begin{figure}
	\begin{center}
		\includegraphics[width=10cm]{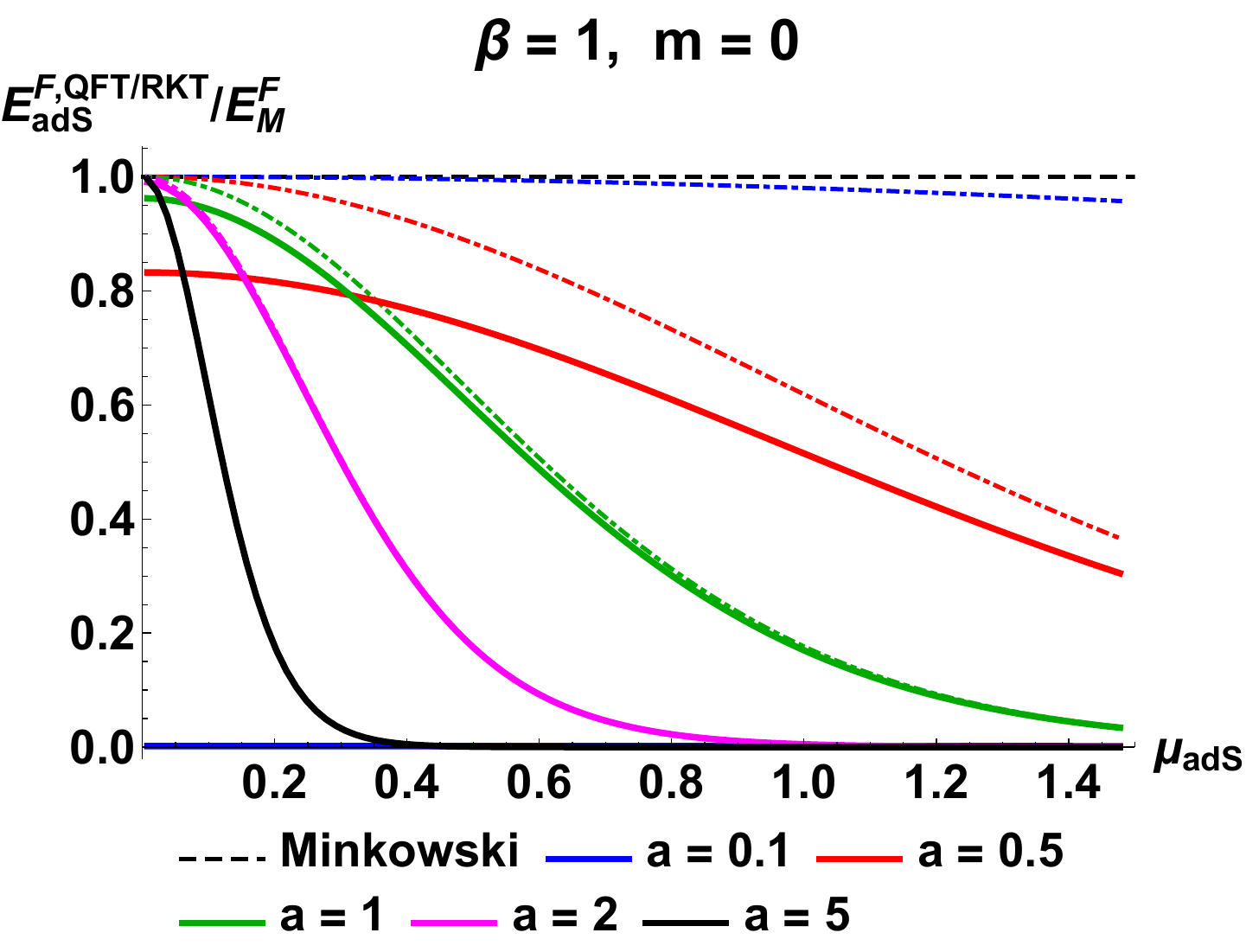}
		\caption{Energy density $E_{\mathrm {adS}}^{F,{\mathrm {QFT/RKT}}}$ divided by the Minkowski space-time energy density $E_{M}^{F}$ (\ref{eq:massless_M}) for massless fermions with inverse temperature $\beta =1$, plotted as a function of the dimensionless geodetic distance $\mu _{\rm {adS}}$ (\ref{eq:muadS}). Solid lines denotes the QFT expressions (\ref{eq:QFTFE}), while dashed lines denote the corresponding RKT quantities (\ref{eq:masslessRKT}). \label{fig:fermiont2}}
	\end{center}
\end{figure}

To see how the energy density depends on the adS radius of curvature $a$, in Fig.~\ref{fig:fermiont2} we plot the RKT and QFT energy densities (\ref{eq:masslessRKT}, \ref{eq:QFTFE}) divided by the Minkowski space-time energy density (\ref{eq:massless_M}) for fixed inverse temperature $\beta =1$ and a selection of values of $a$.  The energy densities are plotted as functions of the dimensionless geodetic distance $\mu _{\rm {adS}}$ (\ref{eq:muadS}).
The overall behaviour of the QFT energy density is similar to that seen in the scalar field case in Fig.~\ref{fig:scalart2}.
 In particular, the QFT energy density at the origin decreases as $a$ decreases; for larger $a$ the energy density profile rapidly decreases to zero as $\mu $ increases, while for smaller $a$ the profile is more spread out in $\mu $. When $a$ is small, the energy density is negligible, being strongly quenched as a result of quantum corrections.
Our use of the dimensionless geodetic distance $\mu _{\rm {adS}}$ (\ref{eq:muadS}) in Fig.~\ref{fig:fermiont2} has an interesting effect on the profiles of the energy density compared to those presented in \cite{Ambrus:2017vlf}, where we used the dimensionful quantity $a\mu _{\rm {adS}}$ as our independent variable. If $\mu _{\rm {adS}}$ is fixed and the adS radius of curvature $a$ increases, then the physical dimensionful geodetic distance also increases.  This is why the profiles in Fig.~\ref{fig:fermiont2} for larger $a$ tend to zero more quickly as $\mu _{\rm {adS}}$ increases than those for smaller $a$. We also see that the profiles become flatter for smaller $a$ when quantum corrections dominate.

\section{Discussion and conclusions}
\label{sec:conc}

In this report we have studied the properties of quantum scalar and fermion fields on four-dimensional adS space-time, considering the v.e.v.s and t.e.v.s of the SET. We began with a brief review of QFT and RKT on Minkowski space-time and discussed the salient features of adS geometry which are important for QFT, in particular the consequences of the fact that  the space-time is not globally hyperbolic.  The maximal symmetry of adS space-time enables us to write the vacuum Feynman Green's function in closed form for both scalar and fermion fields, and to calculate the renormalized v.e.v.~of the SET using Hadamard renormalization. Here we consider the global adS vacuum state, for which the v.e.v.s of the SET preserve the maximal symmetry of adS space-time. 

Making a choice of time, we also define thermal states for scalars and fermions. The fact that we have to choose a time coordinate in order to construct a thermal state breaks the underlying maximal symmetry of adS space-time.  For thermal states, we have restricted our attention to massless fields and refer the reader to \cite{KentEW} and \cite{Ambrus:2017cow} for the extension to massive scalars and fermions respectively. In the scalar case, we have studied fields with conformal and minimal coupling to the Ricci scalar curvature. For fermions, minimal and conformal coupling are the same. We focus on the difference between the t.e.v.~and the v.e.v.~of the SET.  For fermion fields, this takes the perfect fluid form, retaining the symmetry in the space-like directions. However, for scalar fields, this symmetry is also broken (for both conformal and minimal coupling) and the difference between the t.e.v.~and the v.e.v.~of the SET no longer has the perfect fluid form. This breaking of the symmetry between the space-like coordinates can be thought of as arising from  making a choice of origin, relative to which the local inverse temperature (\ref{eq:tolman}) is defined \cite{Allen:1986ty,Panerai:2015xlr}. 
For both scalars and fermions, the radiation accumulates away from the adS boundary, and is concentrated in a region close to the origin.

We have also compared the difference between the t.e.v.~and the v.e.v.~of the SET with the results for the energy density and pressure of a classical thermal gas of particles, calculated in the framework of RKT, to see the effect of quantum corrections.  We find that quantum corrections are more significant for a minimally coupled scalar field than for a conformally coupled scalar field or a fermion field. When the adS radius of curvature is large, the RKT results are a good approximation to the QFT results for fermions and conformally coupled scalars, but not for minimally coupled scalars.  For both scalars and fermions we find that quantum corrections become more significant as the adS radius of curvature decreases and the curvature of the space-time increases.  

\section*{Acknowledgments}

We thank the organizers of the IVth Amazonian Symposium on Physics for the opportunity to present our work and for a very stimulating conference. 
The work of V.E.A.~is supported by a grant from the Romanian National Authority for Scientific Research and Innovation, CNCS-UEFISCDI, project number PN-III-P1-1.1-PD-2016-1423.
The work of E.W.~is supported by the Lancaster-Manchester-Sheffield Consortium for
Fundamental Physics under STFC grant ST/P000800/1 and partially supported by the H2020-MSCA-RISE-2017 Grant No.~FunFiCO-777740.

\end{document}